\documentclass[journal]{IEEEtran} 
\usepackage{amsmath,amsfonts}
\usepackage{algorithmic}
\usepackage{array}
\usepackage[caption=false,font=footnotesize,labelfont=rm,textfont=rm]{subfig}
\usepackage{textcomp}
\usepackage{stfloats}
\usepackage{url}
\usepackage{booktabs} 
\usepackage{siunitx} 
\usepackage{verbatim}
\usepackage{enumerate}
\usepackage{graphicx}
\usepackage{algorithm}
\usepackage{multirow}
\usepackage{amssymb}
\usepackage {float}
\usepackage[switch]{lineno}
\usepackage[backend=biber,style=numeric,sorting=none]{biblatex}
\usepackage{balance}

\begin{document}
\title{
Asymmetric Information Enhanced Mapping Framework for Multirobot Exploration based on Deep Reinforcement Learning}
\author{
	\vskip 1em
	
	Jiyu Cheng, Junhui Fan, Xiaolei Li, Paul L. Rosin, Yibin Li, and Wei Zhang,~\IEEEmembership{Senior Member,~IEEE}
    \thanks{
    This work is supported in part by the Key R\&D Program of Shandong Province under Grants 2025CXGC010111 and 2024CXGC010212, in part by the National Natural Science Joint Foundation of China under Grant U23A20339, and in part by the National Key Laboratory of Human-Machine Hybrid Augmented Intelligence of Xi'an Jiaotong University under Grant HMHAI-202409. \textit{(Corresponding author: Wei Zhang.)}
    
    Jiyu Cheng, Xiaolei Li, and Wei Zhang are with the School of Control Science and Engineering, Shandong University, Shandong, China, 250061, and also with the Key Laboratory of Machine Intelligence and System Control, Ministry of Education, Jinan 250061, China (e-mail: jycheng@sdu.edu.cn; qylxl@sdu.edu.cn; davidzhang@sdu.edu.cn).

    Junhui Fan and Yibin Li are with the School of Control Science and Engineering, Shandong University, Shandong, China, 250061 (e-mail: 202334916@mail.sdu.edu.cn; liyb@sdu.edu.cn).

    Paul L. Rosin is with the School of Computer Science and Informatics, Cardiff University, CF10 3AT Cardiff, U.K. (e-mail: rosinpl@cardiff.ac.uk).   
    
    }
}
\maketitle

\begin{abstract}
Despite significant advancements in multirobot technologies, efficiently and collaboratively exploring an unknown environment remains a major challenge. In this paper, we propose AIM-Mapping, an Asymmetric InforMation enhanced Mapping framework based on deep reinforcement learning. The framework fully leverages the privileged information to help construct the environmental representation as well as the supervised signal in an asymmetric actor-critic training framework.
Specifically, privileged information is used to evaluate exploration performance through an asymmetric feature representation module and a mutual information evaluation module. The decision-making network employs the trained feature encoder to extract structural information of the environment and integrates it with a topological map constructed based on geometric distance. By leveraging this topological map representation, we apply topological graph matching to assign corresponding boundary points to each robot as long-term goal points. We conduct experiments in both iGibson simulation environments and real-world scenarios. The results demonstrate that the proposed method achieves significant performance improvements compared to existing approaches. 

~\\
\end{abstract}
\section{Introduction}
\IEEEPARstart{W}{ith} the advancement of artificial intelligence technology, 
multirobot systems are being increasingly applied in various applications.
In tasks such as search and rescue or inspection \cite{vergnano2012modeling}, robots often operate in unknown environments without prior access to an environment map. During multirobot exploration, robots need to use onboard sensors to perceive and reconstruct the environment efficiently while navigating based on a specific strategy. An early approach, frontier-based exploration \cite{r7}, identifies boundaries of the explored area, guiding robots to ensure complete coverage. These methods typically use geometric information of the robots and frontiers to assign either short-term or long-term goals. As robots move, they continuously update the map to facilitate future decision-making. However, achieving optimal decision-making remains challenging due to factors such as robot mobility, sensor capabilities, and coordination constraints. Some approaches model the environment using occupancy grids or distance-based topological maps, framing the collaborative exploration task as a combinatorial optimization problem \cite{r9,r10,r11}. While optimization solvers can guarantee optimal solutions, they typically suffer from high computational complexity due to the NP-hard property of the problem. To address this challenge, some researchers employ heuristic methods with relaxed constraints to improve target point allocation among robots \cite{yamauchi1997frontier,r13,r14}. These methods, guided by manually designed heuristic functions, offer significant computational efficiency. However, since heuristic rules are based on human intuition, they usually lack generalization ability and often lead to suboptimal and locally constrained decisions.\par

In recent years, deep reinforcement learning (DRL) has achieved significant breakthroughs in solving combinatorial optimization problems \cite{r15} and motion control challenges \cite{r16}. Building on this, multi-agent reinforcement learning (MARL) extends reinforcement learning to the multi-agent domain, demonstrating strong performance in various multirobot applications such as formation control \cite{r17}, autonomous vehicle fleets \cite{r18}, and intelligent warehousing \cite{r19}. Once trained, these strategies enable robots to execute complex coordinated actions. However, challenges remain in exploring unknown environments in both boundary-based and reinforcement learning scenarios: (1) Short-sighted decision-making. Due to the property of the unknown environment tasks, long-term information may be unavailable, making it difficult to determine the information value at the current time step, leading to inaccurate immediate rewards. (2) While multirobot cooperation can significantly improve exploration efficiency, the expanded action space in multirobot scenarios makes finding optimal solutions more complex.\par

In this paper, we propose an efficient multirobot active mapping method called AIM-Mapping. 
During training, privileged information is utilized to mitigate inaccuracies and instability in state value estimation caused by the unknown environment in reinforcement learning. AIM-Mapping encodes partial-map features using a differential structural feature extraction network, generating state values by capturing the difference between privileged information and partial-map. The term “asymmetric” refers to the information asymmetry during reinforcement learning, where the critic module has access to privileged information for unexplored areas that the actor module cannot obtain. The mutual information module serves as an evaluation metric, facilitating policy training. 
Our method, trained on only nine indoor scenes, demonstrates remarkable generalization across different indoor datasets and varying robot numbers. Experimental results highlight its advantage over state-of-the-art multirobot active mapping methods and several adapted reinforcement learning baselines. The main contributions are summarized as follows:
\begin{itemize}
\item[$\bullet$]We propose a novel multirobot active mapping framework of which the collaboration efficiency is greatly enhanced by privileged information based on an asymmetric actor-critic training design.
\item[$\bullet$] We propose a new perspective on evaluating exploration performance in unknown environments, introducing privileged information to assess state value through feature engineering and mutual information.
\item[$\bullet$] We adopt topological graph matching in the multirobot decision-making based on the asymmetric feature representation framework.
\item[$\bullet$] The whole method is deployed and tested in both iGibson simulation environments and real-world scenarios to demonstrate its effectiveness.
\end{itemize}\par
The rest of this article is organized as follows.
Section~\ref{RELATED WORK} introduces some related works as well as their advantages and disadvantages.
Section~\ref{PROBLEM FORMULATION} presents problem formulation of the task.
Section~\ref{Approach} describes the details of our framework.
The experimental implementation and the analysis of the results are presented in Section~\ref{Experimental}.
Finally, Section~\ref{Conclusion} draws the conclusions and proposes our future work.

\section{RELATED WORK}
\label{RELATED WORK}
In this section, we discuss several kinds of multirobot exploration methods, including heuristic methods, optimization-based methods, information-theoretic methods, and learning-based methods.

\textbf{Heuristic methods.} Heuristic methods rely on empirical rules, such as frontier detection or artificial potential fields to guide robots toward unexplored regions. Yamauchi \emph{et al.} \cite{yamauchi1997frontier} proposed the concept of the frontier for active mapping, aiming to guide the robot to the frontiers until the entire space is observed. However, these approaches neglect the effect of coordination, often resulting in redundant exploration. Colares \emph{et al.} \cite{r46} addressed this problem by introducing a collaboration factor to enhance target allocation efficiency. Bourgault \emph{et al.} \cite{r47} utilized Voronoi partitioning to assign robots to frontier points within their subspace, avoiding redundant exploration. The Artificial Potential Field (APF) method creates virtual force fields to guide robots. Initially used in global path planning \cite{r48}, Lau \emph{et al.} \cite{r14} constructed a potential function based on distance to guide movement. However, potential fields based on Euclidean distance usually suffer from local optimum. Renzaglia \emph{et al.} \cite{r49} applied potential fields to local navigation, while Liu \emph{et al.} \cite{r50} designed a nonlinear potential function incorporating coverage factors. More recently, Yu \emph{et al.} \cite{r51} introduced a wave-front distance metric and a penalty function for sensor overlap to reduce redundant exploration and improve efficiency. These methods are simple, computationally efficient, and easy to implement. They perform well in small-scale or structured environments where predefined rules are sufficient. However they often require manual tuning and struggle with scalability in large-scale or complex scenarios.\par
\textbf{Optimization-based methods.} Optimization-based methods formulate exploration as an optimization problem, such as target assignment or path planning, aiming to find optimal or near-optimal solutions\cite{r52}. Werger \emph{et al.} \cite{r9} introduced a cost function based on Voronoi partitioning and used the Hungarian Method for approximate solutions. Klodt \emph{et al.} \cite{r10} proposed a Pairwise Optimization strategy for optimal frontier point allocation. For more complex scenarios, Dong \emph{et al.} \cite{r11} applied a clustering algorithm for frontier points, modeling the problem as an Optimal Mass Transport Problem \cite{r53} with a path distance-based cost function. Faigl \emph{et al.} \cite{r54} modeled target allocation as a multiple Traveling Salesman Problem \cite{r55}. Additionally, Clark \emph{et al.} \cite{r56} modeled the problem as a queue stability control problem, employing Lyapunov optimization \cite{r57} to guide multirobot decision-making.
Wu \emph{et al.} \cite{wu2024bayesian} proposed an adaptive RRT-based frontier detection method and a Bayes-guided evolutionary strategy for scalable multi-robot task allocation, improving efficiency and reducing redundant exploration.
These methods provide theoretically optimal solutions under ideal conditions and can handle multi-objective optimization tasks, such as balancing coverage and energy efficiency.
However, the lack of adaptability in predefined objective functions further restricts their performance in unknown or evolving environments.\par
\textbf{Information-theoretic methods.} 
These methods reduce exploration uncertainty by lowering map entropy and increasing mutual information between sensor data and the environment. They are particularly suitable for environments with high uncertainty or partial observability.
Whaite and Ferrie \emph{et al.} \cite{whaite1997autonomous} introduced strategies for minimizing entropy during exploration of unknown environments, while Elfes \emph{et al.} \cite{elfes1995robot} focused on maximizing mutual information (MI) between sensor data and an occupancy grid map. Information-based exploration strategies aim to minimize uncertainty in robot localization and environment mapping \cite{moorehead2001autonomous,bourgault2002information,visser2008balancing}. In multi-agent systems, \cite{r23,dai2020fast} advanced the idea of using information gain to improve collaboration. In three-dimensional environments, the need for efficient computation of mutual information is particularly pronounced. Henderson \emph{et al.} \cite{henderson2020efficient} proposed a faster method for continuous map computation based on recursive expressions of Shannon mutual information. Asgharivaskasi \emph{et al.} \cite{asgharivaskasi2023semantic} proposed semantic octree mapping and Shannon mutual information computation for robot exploration, deriving an efficiently computable closed-form lower bound for the mutual information between a multiclass octomap and a set of range-category measurements.
Information-theoretic methods are computationally expensive, especially in 3D or large-scale environments. They require accurate probabilistic models, such as occupancy grids, and struggle with scalability due to high communication and computation demands.\par

\textbf{Learning-based methods.} 
Learning-based methods learn exploration policies from environmental feedback, or learn to construct map representation to facilitate decision-making.
Some researchers employed reinforcement learning framework for policy generation.
Geng \emph{et al.} \cite{r58} proposed a decentralized decision-making method in grid map environments using multi-agent reinforcement learning, and robots exchange observation information encoded by convolutional neural networks through a learnable network structure to achieve collaborative decision-making. Later, Geng \emph{et al.} \cite{r59} improved this by introducing attention mechanisms for more targeted information exchange. However, the limited action space in grid maps often results in suboptimal long-term decisions. To address this problem, Tan \emph{et al.} \cite{r60} introduced hierarchical reinforcement learning, extending the decision model to a hierarchical framework. Zhu \emph{et al.} \cite{zhu2024multi} proposed a Two-Stage Coordination (TSC) strategy, which consists of a high-level leader module and a low-level action executor. Researchers have applied reinforcement learning to more realistic scenarios, accounting for robot dynamics and environmental perception. Hu \emph{et al.} \cite{r61} combined DRL with Voronoi segmentation for LiDAR-equipped robots. Chaplot \emph{et al.} \cite{r62} designed a hierarchical framework for single-robot indoor exploration using RGB cameras, while Yu \emph{et al.} \cite{r21} extended this to multirobot vision-based exploration with a Transformer-based decision network. Ye \emph{et al.} \cite{r22} used depth cameras to reconstruct 2D maps and built topological graphs, introducing a multi-path graph neural network to predict distances between boundary points and robots. Lodel \emph{et al.} \cite{lodel2022look} trained an information-aware policy via deep reinforcement learning, that guides a receding-horizon trajectory optimization planner. 
\begin{figure*}
    \centering
    \includegraphics[width=0.8\linewidth]{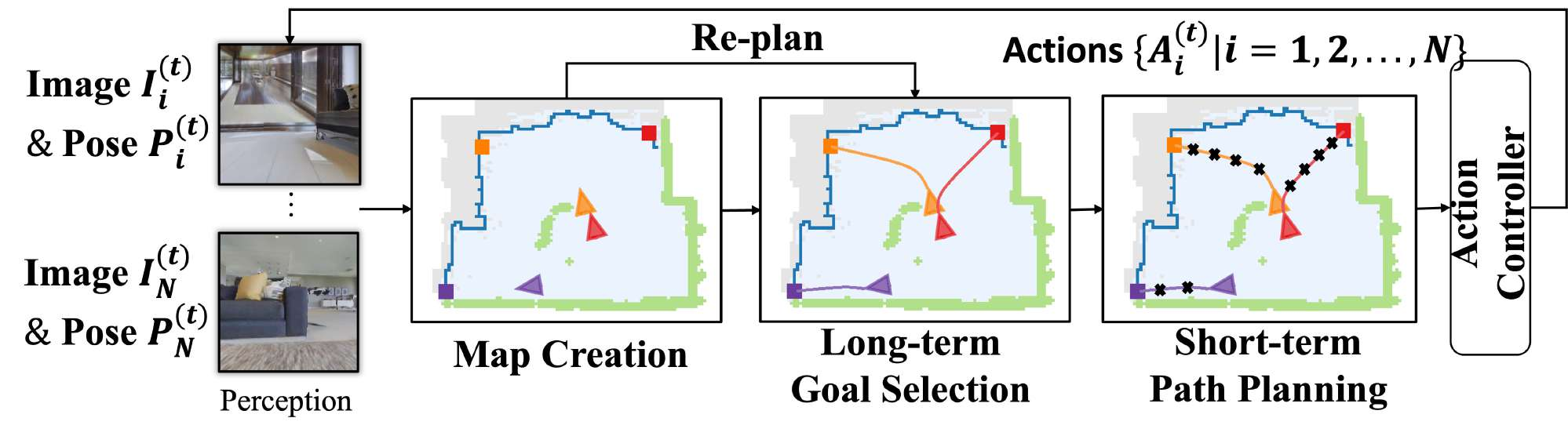}
    \caption{Multirobot collaborative active mapping task contains three main sub-task modules: perception and map creation, long-term goal selection, and short-term path planning.
}
    \label{The overall task framework}
\end{figure*}
Some researchers have also explored the prediction of unexplored areas to facilitate decision-making in robotic exploration.
Shrestha \emph{et al.} \cite{shrestha2019learned} and Katyal \emph{et al.} \cite{katyal2019uncertainty}
proposed to predict the 2D occupancy maps to estimate the total information or
uncertainty for exploration.
Saroya \emph{et al.} 
\cite{saroya2020online} proposed to learn the topological features and used them to inform the exploration policy.
Tao \emph{et al.} \cite{tao2023seer}
detected
frontier clusters, extracted semantic information, and predicted
occupancy and information incrementally during exploration.
In \cite{tao2024learning}, Tao \emph{et al.} proposed an occupancy
prediction module that utilizes the global occupancy map
generated solely from depth images.
Ericson \emph{et al.} \cite{ericson2024beyond} predicted the unseen walls of a partially observed environment to facilitate the robot planning.
Ramakrishnan \emph{et al.} \cite{ramakrishnan2020occupancy} used the robot's egocentric RGB-D observations to infer the occupancy state beyond the visible regions.
Kim \emph{et al.} \cite{kim2023multi} extracted geometric cue from
3D point cloud data and detected the locations of potential cues
such as doors and rooms to help multirobot multiroom exploration.
Zhao \emph{et al.} \cite{zhao2025multirobot} proposed a reinforcement learning-based multirobot exploration strategy that combines dynamic Voronoi partitioning and a multiobjective cost function, enhanced by DDPG and transfer learning for improved adaptability. To enhance the scalability and physical consistency of multi-robot systems, Sebastián \emph{et al.} \cite{sebastian2025physics} have introduced port-Hamiltonian structures combined with attention mechanisms to construct policy networks, enabling sparse and coordinated control within the Soft Actor–Critic framework. Zhu \emph{et al.} \cite{zhu2025autonomous} proposed a hierarchical multi-agent reinforcement learning framework for multi-robot area search, where a role selection module decouples task planning from execution. An intelligent role-switching mechanism enables dynamic transitions between exploration and coverage.

For multirobot indoor exploration, compared with other methods, learning-based paradigm can better utilize high-level feature information, such as the structural information presented in the 2D map. In our method, we adopt the DRL framework and further enhance the mapping performance with asymmetric information. On the one hand, the asymmetric information is used to generate the state value, which depicts the disparity between partial-map and privileged information. On the other hand, asymmetric information helps to formulate the mutual information, which acts as evaluation metric during the training process.  

\section{PROBLEM FORMULATION}
\label{PROBLEM FORMULATION}

\subsection{Multirobot Active Mapping} \par
The multirobot active mapping task can be divided into three sub-task modules: perception and map creation, long-term goal selection, and short-term path planning. In the perception and map creation module, robots transform sensor information into a 2D grid map. In the long-term goal selection module, robots allocate and select long-term goal points on the grid map. In the short-term path planning module, robots plan paths to the selected long-term goal points. The whole task framework is shown in Fig. \ref{The overall task framework}. The policy network and training algorithm focus on long-term goal selection, with existing methods handling perception and map creation, and short-term path planning. \par
\subsubsection{Perception and Map Creation} 
The objective of perception and map creation module is to construct a global map based on sensor information from multiple robots. Robots usually use depth cameras as distance sensors. Multiple robots transform their observation maps into a common world coordinate system based on their positions and orientations, creating a global occupancy map. 
At each time step, the robot obtains depth image $I_i^{(t)}\in \mathbb{R}^{h\times w}$ and the global pose $L_i^{(t)}\in \mathbb{R}^3$ from the environment. 
The perception and map creation module aims to construct a top-down 2D map to describe environmental features. 
The occupancy grid map of robot \( i \) at time \( t \) is denoted as \( O_i^{(t)} \in \{0,1\}^{X_l \times Y_l \times 2} \), where \( X_l \) and \( Y_l \) represent the predefined map dimensions. 
In this study, only obstacles that hinder the robot’s movement are considered during mapping. 
At each time step $t$, multiple robots collaboratively update the global observation $O^{(t)}$ through exchanging pose estimates $L_i^{(t)}=(x,y,\theta)$ and local maps $O_i^{(t)}$, aligning these maps to the world coordinate system and integrating them with the prior global observation $O^{(t-1)}$.

\subsubsection{Long-term Goal Selection} 
In long-term goal selection, a straightforward approach for each robot is to move toward the boundaries of the explored area at each time step. Assuming that the map is closed and bounded, as long as robots continuously move toward boundary points, they will eventually complete exploration of the environment. This method based on boundary points is also adopted in this paper. Therefore, the goal is to assign a boundary point as a long-term target point for each robot when each planning cycle arrives, allowing multiple robots to explore as much of the unknown environment as possible in the shortest possible time and ultimately establish a global grid map containing all environmental information. \par
\subsubsection{Short-term Path Planning} \par
Short-term path planning is a discretized subtask in which robots, after receiving long-term goals, individually plan the shortest paths to reach their respective targets based on the global grid map $O^{(t)}$. In this paper, the Fast Marching Algorithm \cite{r88} is adopted to compute the shortest path from the robot to the target position. Upon obtaining the short-term path points, robots generate low-level actions through a simple heuristic method \cite{r87}: if a robot is facing the path point, it executes a forward action; otherwise, it performs rotation actions until it faces the path point.\par
\subsection{Modeling of Markov Decision Process}\par
The long-term goal selection problem constitutes a critical component in multi-robot active mapping tasks. For indoor exploration scenarios, this cooperative mapping task can be formally modeled as a centralized Partially Observable Markov Decision Process (POMDP), defined by the tuple $\langle \mathcal{N}, \mathcal{S}, \mathcal{O}, O, \mathcal{A}, \mathcal{P}, \mathcal{R}, \gamma \rangle$. Where $\mathcal{N}$ is the set of N agents. $\mathcal{S}$ represents the global state space. $\mathcal{O}=\times_{i\in \mathcal{N}}\mathcal{O}^i$ is the joint observation space for multiple agents, where $O$ is the observation function. 
$\mathcal{A}=\times_{i\in \mathcal{N}}\mathcal{A}^i$, represents the joint action space for multiple agents. $\mathcal{P}:\mathcal{S}\times \mathcal{A}\rightarrow\triangle(\mathcal{S})$ denotes the state transition probabilities. $\mathcal{R}:\mathcal{S}\times\mathcal{A}\rightarrow\mathbb{R}$, is the reward function for all agents. $\gamma\in [0,1)$ is the reward discount factor. At each time step, agents receive local observations $o^{(t)}_i=O(s_t,i)$ from the global state $s_t\in\mathcal{S}$ then a controller collects observations from all agents and generates a joint action $a_t=\pi (a_t|o_1^{(t)},...,o_N^{(t)})$ through a centralized policy. Each agent receives and executes the corresponding action $a^i_t\in a_t$ from the central controller. Finally, the joint actions $a_t \in \mathcal{A}$ of multiple agents transition the system from state $s_t$ to state $s_{t+1}$ based on the state transition probabilities $P(s_{t+1}|s_t,a_t)$ and receive reward $r^{(t)}=R(s_t,a_t)$. 
To address this problem, value networks and policy networks are designed, employing an end-to-end multi-agent deep reinforcement learning framework. The agent starts from state $s$, follows the policy $\pi$ to make decisions, and the expected return is used to evaluate the quality of the policy. This framework aims to maximize the state-value function $V_\pi (s)=\mathbb{E}_{s,a}[\sum_{t=0}^T \gamma^{(t)} r^{(t)}|s_0=s, a_t \sim  \pi (\cdot |o^{(t)}_1,\dots,o^{(t)}_N)]$ to learn an optimal centralized policy $\pi^* (\cdot |o^{(t)}_1,...,o^{(t)}_N)$, where $T$ denotes the total number of training rounds. In this task, each agent's action space consists of a set of candidate points for long-term goal selection. The agent selects a candidate point as its long-term goal and moves toward it.
\begin{figure*}
    \centering
    \includegraphics[width=1\linewidth]{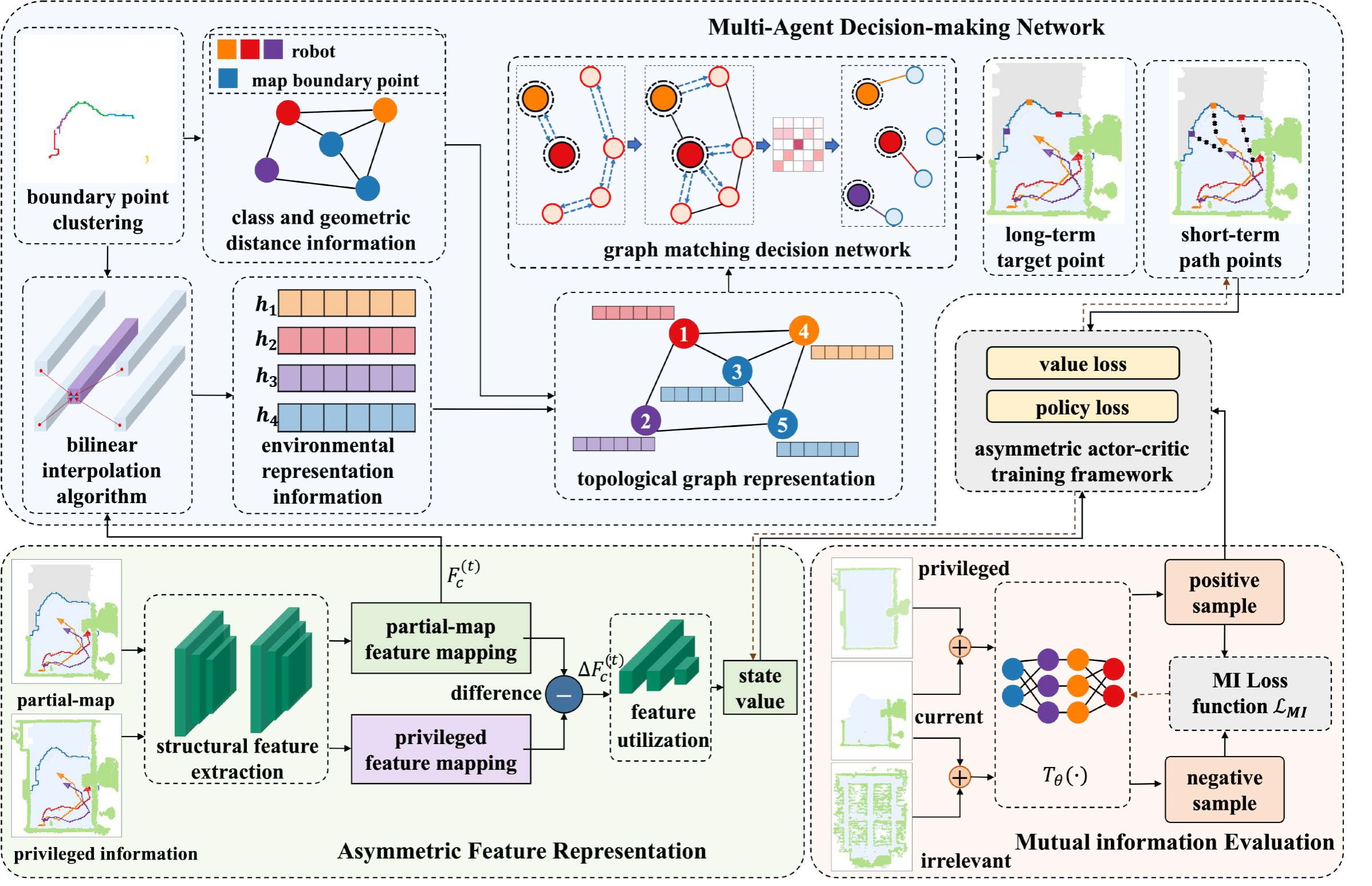}
    \caption{The overall AIM-Mapping framework. Asymmetric Feature Representation is used to generate the state value and the partial-map feature mapping. Multi-Agent Decision-making Network combines the geometric distance information and structural information to formulate the topological graph representation, and adopts graph matching to generate the corresponding goal point. Mutual Information Evaluation is utilized to facilitate the training process. Solid black arrows represent the forward data flow through the network. Brown dashed arrows indicate the gradient backpropagation paths used during training for updating network parameters.
}
    \label{ATR-Mapping overall frame diagram}
\end{figure*}
\section{METHODOLOGY}
\label{Approach}
This paper introduces AIM-Mapping, a novel deep reinforcement learning framework for collaborative multirobot exploration. The core idea is to fully utilize the privileged information to help enhancing the efficiency of multirobot exploration. There are mainly three modules: \emph{Asymmetric Feature Representation} (AFR), \emph{Mutual Information Evaluation} (MIE), and \emph{Multi-Agent Decision-making Network} (MADN).\par
During the training process, the AFR module leverages both partial-map data and privileged information to generate state values and partial-map feature mappings. The MADN module then utilizes topological information—derived from geometric distances—together with structural features extracted from the partial-map feature mappings to construct a topological graph representation of the environment. Through graph matching, each robot is assigned a corresponding boundary point as its long-term goal. Subsequently, the robot generates short-term target points for navigation using a path planning algorithm. MIE module is specially designed to assist the training process by using privileged information to quantify information gain and the reduction of measurement uncertainty during exploration. 
During the testing phase, only partial-map data is fed into the trained feature extractor from the AFR module. The resulting partial-map feature mapping is then passed to the MADN module, which derives optimal clusters of long-term goal points based on the established topological structure. Finally, robots perform short-term path planning to reach their assigned goals. The overall AIM-Mapping framework is illustrated in Fig.\ref{ATR-Mapping overall frame diagram}.

\subsection {Asymmetric Feature Representation} \par
This section provides an overview of the Asymmetric Feature Representation, which is designed to encode the disparity between partial-map and privileged information.
As shown in Fig. \ref{ATR-Mapping overall frame diagram}, the partial-map and privileged information act as the input.
In order to maintain the structural properties, both types of information are represented using a grid map.
The grid map consists of 5 channels, including an obstacle channel, a passable-area channel, a robot channel, a boundary channel, and a trajectory channel.
Each channel is encoded with a binary map, where the value of a grid cell is 1 if there is corresponding entity information, otherwise the value is set as 0. Specifically, the partial-map is represented as $m_c^{(t)}\in \big\{ 0,1 \big\}^{X\times Y\times5}$, where $X$ and $Y$ are the dimensions of the global map. And the input privileged information is represented as $\hat m_c^{(t)}\in \big\{ 0,1 \big\}^{X\times Y\times5}$. 
A feature encoding network is designed to capture structural-spatial correlations, implemented as a Structural Convolutional Network (SCNet) with hierarchical downsampling. The SCNet backbone transforms $X\times Y\times5$ inputs into compact latent representations $X_h\times Y_h\times C_h$  ($X_h=X/8$, $Y_h=Y/8$, $C_h=32$), where spatial structures are encoded as channel-wise activation patterns.
At time $t$, the privileged information and partial-map, after feature extraction by the SCNet, yield privileged feature mappings $\hat F_c^{(t)}\in \mathbb{R}^{X_h\times Y_h\times C_h}$ and partial-map feature mappings $F_c^{(t)}\in \mathbb{R}^{X_h\times Y_h\times C_h}$, respectively. The sizes of privileged feature mappings and partial-map feature mappings are the same, differing only in whether the channel information representing explored areas includes privileged information.As a result, the disparity between partial-map feature mapping and privileged feature mapping captures the difference between explored area and the whole area, as well as the structural information. 

Specifically, the differential calculation process described above can be represented as:\par
\begin{equation}
F_c^{(t)}=\mathrm{SCNet}\big( O_c^{(t)}\big) ,\hat F_c^{(t)}=\mathrm{SCNet} \big( \hat O_c^{(t)}\big)
\end{equation}
\begin{equation}
\Delta F_c^{(t)}=Flatten \big( F_c^{(t)}-\hat F_c^{(t)} \big)
\end{equation}

\begin{equation}
state\_value=\mathrm{FUNet}(\Delta F_c^{(t)})
\end{equation}
Where $\Delta F_c^{(t)}$ is the vectorized disparity between feature mappings. It is worth noting that in the above equation, the $\mathrm{SCNet}$ network shares parameters. %
By the processing of the network structure in the terms of differences, the feature encoding network can be enhanced, thereby encoding key structural information of the environment into the feature mappings. \par
In the proposed method, the policy network takes $F_c^{(t)}$ as input, while the value network receives $\Delta F_c^{(t)}$.
These features are processed by the lower-level actor-critic networks, where $F_c^{(t)}$ directly supports decision-making, and
$\Delta F_c^{(t)}$ enhances training performance. 
During training, both $F_c^{(t)}$ and $\Delta F_c^{(t)}$
are utilized to optimize the structural feature extractor. However, during deployment, only $F_c^{(t)}$ is required for inference.
The asymmetric feature representation is supervised implicitly. As a result, $\Delta F_c^{(t)}$ is only used in the training process. Despite the implicit supervision of feature representation, the more accurate feature extraction will promote more rational robot actions, leading to higher episode rewards. These episode rewards are then used by the value network to perform backpropagation, supervising the training of the feature extraction network. When the network is fully trained, the structural feature extractor will be optimized, which means $F_c^{(t)}$ encodes well the structural information of the environment. During deployment, the policy network receives $F_c^{(t)}$, and conducts decision-making based on the encoded information of $F_c^{(t)}$.\par

\subsection{Mutual Information Evaluation}\par
To quantify information gain and reduction in measurement uncertainty during exploration, we design a mutual information evaluation network and propose a mutual information estimation framework based on variational inference. This framework leverages a neural network to implicitly model the dependency between the global map $M$ and the joint local partial-map of multiple robots $m^{(t)}$, thereby providing efficient exploration reward signals for reinforcement learning. Here, the global map $M$ and the local partial-map $m^{(t)}$ only contain information regarding obstacles and passable-area. Mutual information is defined as
\begin{equation}
I(M; m^{(t)}) = H(M) - H(M|m^{(t)})
\end{equation}
where $H(M)$ is the Shannon entropy quantifying prior map uncertainty, and $H(M|m^{(t)})$ measures the residual uncertainty after incorporating partial-map $m^{(t)}$. Direct computation of this Kullback-Leibler (KL) divergence is intractable in high-dimensional spaces, motivating our variational approach.
\begin{equation}
\begin{aligned}
I(M; m^{(t)})
= -\mathbb{E}_{P(M)}[\log P(M)]\\
+\mathbb{E}_{P(M,m^{(t)})}[\log P(M|m^{(t)})]
\end{aligned}
\end{equation}
Here, $P(M)$ denotes the prior probability distribution of the global map, encoding the initial uncertainty about the environment. The conditional probability $P(M|m^{(t)})$ captures how this uncertainty is reduced upon observing $m^{(t)}$. 
From probability theory, the mutual information can be equivalently expressed as the KL-divergence between the joint distribution $P(M,m^{(t)})$ and the product of marginals $P(M)P(m^{(t)})$:
\begin{equation}
I(M; m^{(t)}) = D_{\mathrm{KL}}(P(M, m^{(t)}) \parallel P(M)P(m^{(t)}))
\end{equation}
Using the Donsker-Varadhan duality, we establish a tractable lower bound:
\begin{equation}
\begin{aligned}
I(M; m^{(t)}) \geq \sup_{T \in \mathcal{T}} \mathbb{E}_{P(M, m^{(t)})}\left[T(M, m^{(t)})\right] \\
- \log\,\mathbb{E}_{P(M)P(m^{(t)})}\left[e^{T(M, m^{(t)})}\right]
\end{aligned}
\end{equation}
where $\mathcal{T}$ is a class of functions that satisfy appropriate integrability conditions. We parameterize the function $T$ using a neural network $T_\theta$, which enables us to approximate complex functions without explicit density estimation and to perform end-to-end training.\par
To approximate the expectation terms in the variational lower bound during training, we introduce the mechanism of positive and negative sample pairs:
\begin{itemize}
    \item \textbf{Positive Sample Pairs:} Positive sample pairs are directly drawn from the joint distribution $P(M, m^{(t)})$, reflecting the true dependency between the global map and the local partial-map. In practice, when an agent takes an action in the environment, it simultaneously obtains the local partial-map $m^{(t)}$ and the corresponding global map information $M$, forming a positive sample pair $(M, m^{(t)})$.
    \item \textbf{Negative Sample Pairs:} To approximate the product of the marginal distributions $P(M)P(m^{(t)})$, we construct negative sample pairs by disrupting the correspondence in the positive pairs. Specifically, for a given positive sample pair $(M, m^{(t)})$, we randomly select a global map $\hat M$ from other positive pairs (ensuring that $\hat M$ is independent of the current $m^{(t)}$) to form a negative sample pair $(\hat M, m^{(t)})$. This shuffling breaks the original dependency between $M$ and $m^{(t)}$, effectively simulating independent sampling.
\end{itemize}\par
\begin{figure*}
    \centering
    \includegraphics[width=1\linewidth]{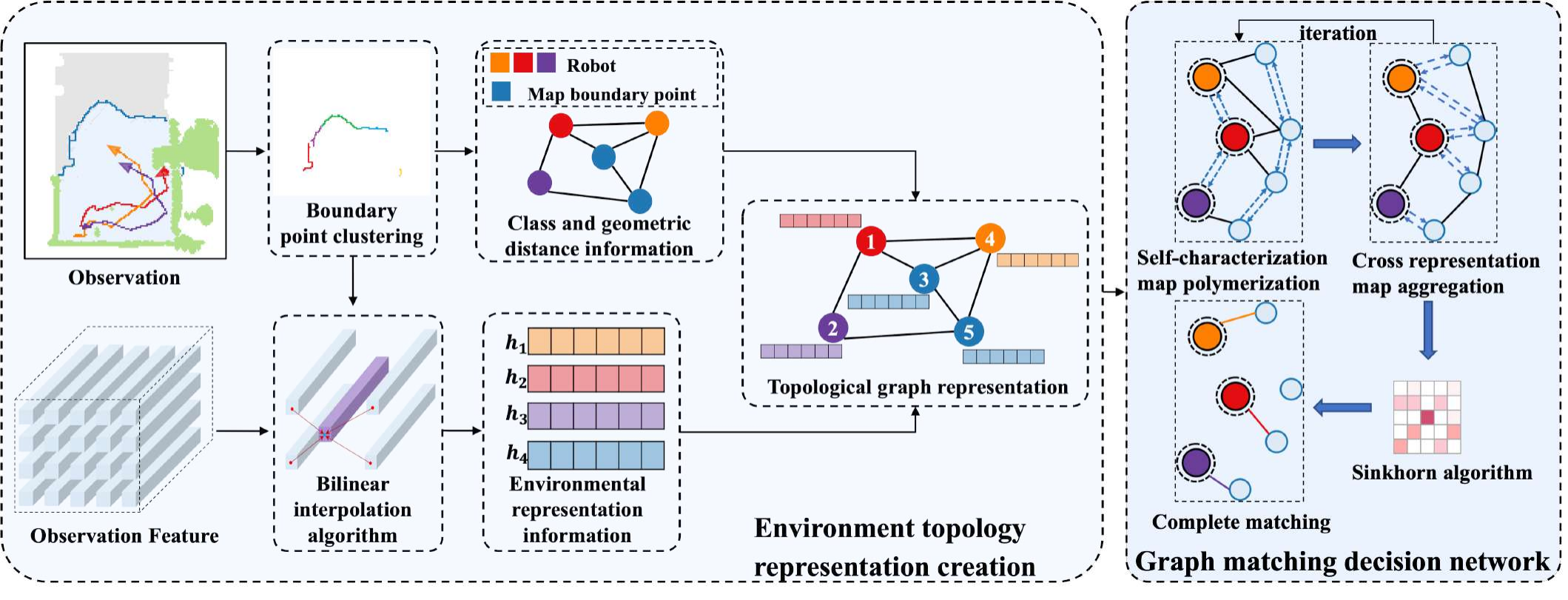}
    \caption{Multi-agent decision-making network based on topological graph matching. This framework concatenates internal and external information fusion of the graph, completing graph matching between the representation of robots and boundary points, and assigning corresponding boundary points as long-term target points for each robot.
}
    \label{Multi-agent decision-making network based on topological graph matching}
\end{figure*}
In training, we use mini-batches of positive and negative sample pairs to approximate the two expectation terms in the variational lower bound. Suppose a mini-batch contains $N$ samples. The expectation over the positive sample pairs is approximated as:
\begin{equation}
\mathbb{E}_{P(M, m^{(t)})}\left[T(M, m^{(t)})\right]\approx \frac{1}{N} \sum_{i=1}^{N} T_\theta(M, m^{(t)}_i)
\end{equation}
For the negative sample pairs, we approximate this expectation using all possible combinations (or a subset of non-corresponding pairs, i.e., $i \neq j$) in the mini-batch:
\begin{equation}
\begin{aligned}
\log  \mathbb{E}_{p(M)p(m^{(t)})}\left[e^{T(M, m^{(t)})}\right] \\
\approx \log \left( \frac{1}{N^2} \sum_{i,j=1}^{N} e^{T_\theta(\hat M_j, m^{(t)}_i)} \right)
\end{aligned}
\end{equation}
The loss function is then defined as the negative of the variational lower bound: 
\begin{equation}
\begin{aligned}
\mathcal{L}_{MI}(\theta) = -\frac{1}{N} \sum_{i=1}^{N} T_\theta(M, m_i^{(t)})\\ 
+ \log \left( \frac{1}{N^2} \sum_{i,j=1}^{N} e^{T_\theta(\hat M_j, m_i^{(t)})} \right) 
\end{aligned}
\end{equation}
By optimizing the neural network $T_\theta$ through backpropagation, we minimize the loss function $\mathcal{L}_{MI}(\theta)$, which effectively tightens the variational lower bound of mutual information. This process provides more accurate and efficient exploration rewards for reinforcement learning, as the estimated mutual information directly reflects the reduction in environmental uncertainty achieved through agent exploration. These properties make our method particularly suitable for multi-agent reinforcement learning, where efficient and stable exploration is critical for coordinating agents in large-scale, uncertain environments.

\subsection {Multi-Agent Decision-making Network} \par
To maximize the utilization of environmental structural information and enable multirobot systems to make rational decisions, a topology-based graph matching decision network is designed. First, feature vectors corresponding to robot positions and boundary points in the partial-map feature map are extracted and combined with geometric distance features from the environment to construct a representation of the topological graph. Then, a graph neural network framework is adopted to perform graph matching, using the design from \cite{r22} as the network backbone. The overall framework process is illustrated in Fig. \ref{Multi-agent decision-making network based on topological graph matching}.\par

\subsubsection {Point Feature Extraction}\par
After obtaining feature mapping from the Asymmetric Feature Representation module, which encodes the structural information of the environment, a point feature extraction method based on nearest neighbor clustering and bilinear interpolation algorithm is adopted. The method first clusters the boundary points of the explored area, using the centroid of each cluster as the representative of corresponding boundary point cluster, and also records the number of points in each cluster as one of the inputs for decision-making. Additionally, the maximum distance between two boundary points in the same cluster cannot exceed a threshold distance $r_{clus}$. After clustering is completed at time $t$, the boundary point cluster $i$ can be represented as $F_i^{(t)}=\{ f_k^{(t)} \}_{k=1:n_i}$ where $f_k^{(t)}\in \mathbb{R}^2$ represents the two-dimension coordinates of boundary point $K$, and $n_i$ indicates the number of boundary points in cluster $i$. And the centroid of boundary point cluster $F_i^{(t)}$ can be represented as $f_i^c= \frac{1}{n_i} \sum_{k=1}^{n_i} f_k^{(t)}$. The specific clustering algorithm and computational process are illustrated in the pseudocode of Algorithm 1.\par
Extracting the structural features of key locations such as robots and boundary points is essential for the decision-making network. In our method, bilinear interpolation is applied to the feature map to extract the relevant feature vectors corresponding to the original grid coordinates in the partial-map feature map. 
Given an input partial-map with the size of $X\times Y\times 5$ and a feature map with the size of $X_l\times Y_l\times C_h$, for a point $p_i$ with coordinates $(x,y)$ in the input partial-map, its projected coordinates in the feature map space should be $p'_i=(x',y')$, where $x'=x\cdot \frac{X_l}{X},y'=y\cdot \frac{Y_l}{Y}$. The value corresponding to point $(x',y')$ in the feature map $F_c^{(t)}\in \mathbb{R}^{X_l\times Y_l\times C_h}$ can be obtained using bilinear interpolation along the x-axis and y-axis directions. 
Assume that the feature points closest to the distance feature mapping point $(x',y')$ are $(x_0,y_0)$, $(x_0,y_1)$, $(x_1,y_0)$ and $(x_1,y_1)$ with feature vector boundary values $F^{(t)}_c(x_0,y_0)$, $F^{(t)}_c(x_0,y_1)$, $F^{(t)}_c(x_1,y_0)$ and $F^{(t)}_c(x_1,y_1)$. The feature vector $I_i^f$ corresponding to the point $p_i=(x,y)$ in the partial-map feature mapping can be expressed as
\begin{equation}
\begin{aligned}
I_i^f=(x_1-x') \cdot (y_1-y') \cdot F^{(t)}_c(x_0,y_0) \\
+ (x_1-x') \cdot (y'-y_0) \cdot F^{(t)}_c(x_0,y_1) \\
+ (x'-x_0) \cdot (y_1-y') \cdot F^{(t)}_c(x_1,y_0) \\
+ (x'-x_0) \cdot (y'-y_0) \cdot F^{(t)}_c(x_1,y_1)
\end{aligned}
\end{equation}
The bilinear interpolation process described above is denoted as $I_i^f=Interp(p_i,F_c^{(t)})$ in this paper.\par

\begin{algorithm}
    \renewcommand{\algorithmicrequire}{\textbf{Input:}}
    \renewcommand{\algorithmicensure}{\textbf{Output:}}
    \caption{Adjacent Neighbor Clustering of Boundary Points}
        \begin{algorithmic}[1]
            \REQUIRE  A set of boundary points $F=\{ f_k \}_{k=1:n}$ containing $n$ boundary point.
            \ENSURE The boundary point clusters \( n_c \) and their centroids are represented as \( F_{cluster} = \{ F_i, f_i^c \}_{i=1:n_c} \), where \( F_i = \{ f_j \}_{j=1:n_i} \).

            \STATE Initializes the clustering of boundary points $F_{cluster}$;
            \WHILE{\textbf{not} $F=\varnothing$ }
                \STATE Initialize a cluster, called $F_i$;
                \STATE Take a boundary point from the boundary point set $F$ and add it to the cluster $F_i$;
                \STATE \textbf{repeat}
                    \FOR{$f$ in $F$}
                        \IF{$f$ is any neighboring boundary point of a cluster, and the distance to any boundary point within cluster $F_i$ is less than $r_{clus}$}
                            \STATE Remove $f$ from the set $F$ and add it to the cluster $F_i$;
                        \ENDIF
                    \ENDFOR

            \STATE \textbf{until} No new boundary points are added to the cluster $F_i$;
                \STATE Compute the average distance between each boundary point in the cluster $F_i$ and the remaining boundary points in the cluster. Select the boundary point with the smallest average distance as the cluster center point$f_i^c$;
                \STATE Add the cluster $F_i$ and the cluster center point $f_i^c$ to the set of boundary point clusters $F_{cluster}$;
            \ENDWHILE
        \end{algorithmic}
\end{algorithm}
\subsubsection{Topological Graph Representation}
After point feature extraction, the node features need to be combined with geometric distance information from the environment to construct self-representation graphs $G_r=\{V_r,E_r\}$ and $G_f=\{V_f,E_f\}$, which only contain information about robots or boundary points, as well as a cross-representation graph $G_r=\{V_r,V_f,E_{rf}\}$, which contains both robot and boundary point information. For a node $i$ in the self-representation or cross-representation graph, its initial node feature vector $v_i\in \mathbb{R}^{5+C_h}$ is composed of three parts: category information $v_i^{cla}\in \{0,1\}^2$, geometric information $v_i^{geo}\in \mathbb{R}^3$, and environmental representation information $v_i^{rep}\in \mathbb{R}^{C_h}$. The category information $v_i^{cla}$ is a one-hot encoded label indicating whether the node is a robot node or a boundary point. The first two dimensions of geometric information $v_i^{geo}$ represent position $p_i=(x,y)$ of the node, while the last dimension represents the geometric information of the node. Additionally, the environmental representation information is obtained by interpolating the node position information in the partial-map feature map: $v_i^{rep}=Interp(p_i,F_c^{(t)})$.\par
The environment is partially observable, and utilizing historical observation information helps robots avoid redundant exploration. The self-representation graph $G_r^h=\{V_r^h,E_r^h\}$ is constructed to represent the historical trajectory information of robots, and $G_g^h=\{V_g^h,E_g^h\}$ is constructed to represent historical boundary point information. To establish connections between current robot and boundary point information and historical information, we construct the cross-representation graph $G_{r\_r}^h=\{V_r^h,V_r,E_{r\_r}^h\}$, as well as the cross-representation graph $G_{f\_g}^h=\{V_g^h,V_f,E_{f\_g}^h\}$ between current boundary points and historical target points. By constructing these topological graphs, both structural feature information and geometric distance information in the environment can be adequately represented, laying the foundation for efficient decision-making by subsequent graph matching decision networks.

\subsubsection {Graph Matching Decision Network}
Graph matching decision network utilizes a graph attention mechanism to sequentially aggregate and extract features from the self-representation graph and the cross-representation graph, updating the features of corresponding edges and nodes in the topological graph. In the cross-representation graph $G_{rf}$ after feature updates, the feature value of each edge in the edge set $E_{rf}$ represents the matching degree of each robot to the boundary point node. Therefore, by extracting the feature values from the updated edge $E_{rf}$ set and using the Sinkhorn algorithm for linear assignment computation, the graph matching can be completed, assigning long-term target points to each robot.\par

We used an encoder based on a multi-layer perceptron network to encode the category information and geometric information $[v_i^{cla},v_i^{geo}]$ of each node $i$ in the graph, obtaining a feature vector of length $C_h$. This feature vector is concatenated with the environmental representation information $v_i^{rep}$ to form the node feature $v_i^0\in \mathbb{R}^{2C_h}$ for subsequent feature aggregation. The core idea of the graph attention network is to utilize an attention mechanism to aggregate features between neighboring nodes in the topological graph. Therefore, for the nodes $v_i^l$ in the $l-th$ layer of the graph network, trainable weight parameters $W_k^l$, $W_q^l$, and $W_v^l$ are introduced to generate the key $k_i^l$, query $q_i^l$, and value $u_i^l$ in the attention mechanism:\par
\begin{equation}
k_i^l=W_k^l\cdot v_i^l,q_i^l=W_q^l\cdot v_i^l,u_i^l=W_v^l\cdot v_i^l
\end{equation}\par
The attention coefficient $a_{ij}^l$ between node $i$ and its neighboring node $j\in N_i$ can be calculated by the following equation: \par
\begin{equation}
a_{ij}^{l,self}=\frac{exp(k_j^{l^T}\cdot q_i^l)}{\sum_{m\in N_i}exp(k_m^{l^T}\cdot q_i^l)}
\end{equation}\par
Additionally, the attention coefficient $a_{ij}^l$ will also serve as the edge feature value between node $i$ and its neighboring node $j$ in the topological graph. Therefore, for node $i$, the aggregation of neighboring node feature values $v_{N_i}^l$ can be expressed as:
\begin{equation}
v_{N_i}^l=\sum_{m\in N_i}a_{im}^l\cdot v_j^l
\end{equation}\par
The final feature value of node $i$ will be updated as the aggregation of neighboring node features and the fusion with its own node feature:
\begin{equation}
v_i^{l+1}=v_i^l+\rho ([v_i^l||v_{N_i}^l])
\end{equation}\\
Where $\rho (\cdot )$ represents the feature fusion function implemented using a multilayer perceptron, and $[\cdot ||\cdot ]$ denotes the concatenation of two feature values. The feature of each node in the self-representation graph will be updated to the aggregation of its own feature and the features of its neighboring nodes. This operation will be applied to the self-representation graphs $G_r$, $G_f$, $G_r^h$, and $G_g^h$ in this method. After completing the feature updates in the self-representation graphs, the node features will serve as the initial features for the corresponding nodes in the cross-representation graph, which will then be input into the subsequent graph attention network. Unlike that in the self-representation graphs, the feature extraction process in the cross-representation graph uses a nonlinear mapping method to generate attention coefficients. Additionally, the distance $d_{ij}$ calculated by the fast marching algorithm is incorporated as an input to the non-linear mapping function $\varphi (\cdot )$:\par
\begin{equation}
a_{ij}^{l,cross}=\frac{exp(\varphi ([k_j^l||q_i^l||d_{ij}]))}{\sum_{m\in N_i}exp(\varphi ([k_m^l||q_i^l||d_{im}]))}
\end{equation}\par
The notation $[\cdot ||\cdot|| \cdot ]$ denotes the concatenation of three vectors. Additionally, the non-linear mapping function $\varphi (\cdot )$ is implemented using a multilayer perceptron, which outputs a one-dimensional real number. In this study, the cross-representation graphs $G_{r\_r}^h$, $G_{f\_g}^h$, and $G_{rf}$ will undergo sequential feature extraction using graph attention networks. The edge features in the cross-representation graph $G_{rf}$ will be extracted and used as the affinity matrix in graph matching, denoted as $A_M\in \mathbb{R}^{n_r\times n_f}$. Here, $n_r$ and $n_f$ represent the numbers of robots and boundary points, respectively. Each element in the matrix represents the degree of matching between the corresponding robot and boundary point. Ultimately, we employ the Sinkhorn algorithm to iteratively normalize the rows and columns of the affinity matrix alternately, gradually transforming it into a probability matrix to accomplish graph matching. Each robot will select the boundary point with the highest probability value from the probability matrix as its long-term target point.\par

\subsection{Asymmetric Actor-Critic Training Framework}\par
We jointly optimize the policy network (Multi-Agent Decision-making Network) with Asymmetric Feature Representation via reinforcement learning to maximize cumulative rewards over the entire task horizon. In multi-robot environments, our global planner’s centralized decision-making architecture enables the use of the off-policy Proximal Policy Optimization (PPO) \cite{r75} algorithm for policy optimization.  

The objective of active mapping is to achieve high time efficiency and map completeness. To this end, we design a temporal reward \( R_{\text{time}} \) and a coverage reward \( R_{\text{coverage}} \). The temporal reward encourages efficient exploration by penalizing unnecessary time steps, formally defined as:  
\begin{equation}  
R_{\text{time}} = -0.01  
\end{equation}  
If the cumulative area explored by the robot team at decision step \( t \) is denoted as \( A^{(t)} \), the coverage reward \( R_{\text{coverage}} \) is defined as the incremental coverage area (in \( \mathrm{m}^2 \)):  
\begin{equation}  
R_{\text{coverage}}^{(t)} = A^{(t)} -A^{(t-1)}
\end{equation}  \par 
To ensure training stability, we introduce a dynamically adjusted weighting parameter \( d_t \) that decays with the number of training steps. A typical implementation defines \( d(t) \) as: $d_t = d_0 \cdot e^{-\lambda t}$ with \( d_0 \) being the initial weight and \( \lambda \) controlling the decay rate. 
 This parameter balances the exploration rate (via coverage reward) and the mutual information loss during training, enabling a gradual transition from coverage-driven exploration to mutual information optimization.  
\begin{equation}  
R^{(t)} = R_{\text{time}} + d_t \cdot R_{\text{coverage}}^{(t)}
\end{equation}  \par 
During training, multiple agents interact with the environment following the policy \( \pi_\theta \), collecting trajectories of state-action-reward tuples into a shared replay buffer. At the end of each episode, the algorithm samples a batch of \( N_b \) transitions from the buffer for training. The loss function of the PPO algorithm is formulated as:  
\begin{equation}  
\mathcal{L}_{\text{PPO}} =\mathcal{L}_{\text{policy}}+ c_1 \cdot \mathcal{L}_{\text{value}} - c_2 \cdot \mathcal{L}_{\text{entropy}} + c_3\cdot \mathcal{L}_{\text{MI}}
\end{equation}  
Where $\mathcal{L}_{\text{policy}}$ denotes the policy gradient loss, $\mathcal{L}_{\text{value}}$ represents the value function loss, $\mathcal{L}_{\text{entropy}}$ is the entropy regularization term to encourage exploration, $c_1$, $c_2$ and $c_3$ are weighting coefficients. \par
To account for the increasing reliability of mutual information estimation during training, the weight $c_3$ follows an exponential curriculum schedule: $c_3=\hat{c}_3 \cdot e^{kt}$ where $\hat{c}_3$ is the initial scaling factor and $k$ controls the growth rate. This design ensures that the mutual information objective gains influence progressively as its estimation becomes more accurate.

\section{Experiments}
\label{Experimental}
\subsection{Experimental  Setup} \par
\subsubsection{Experimental Environment}\par
To validate the proposed framework, experiments were conducted using the iGibson physics simulation engine. iGibson is a virtual environment tool for robotics and AI research, providing realistic indoor scenes for the development and testing of robot perception, navigation, and task planning. The iGibson simulation engine supports various map scene datasets and realistic physics-based interactions between robots and environments. In these experiments, TurtleBot robots equipped with depth cameras were used within the iGibson simulation engine to closely simulate real-world scenarios. The TurtleBot robots can move using a differential drive method and perceive the environment through depth cameras, with realistic collision interactions with the environment. Note that in simulated experiments, we assume that sensors are noise-free, which means that at each time step, robots can get accurate localization and generate accurate partial-maps.\par
For the experiments, publicly available Gibson and MatterPort3D datasets were used for training and testing, respectively. The Gibson dataset offers large-scale 3D data of real indoor environments, while the MatterPort3D dataset provides a larger scale and more diverse set of indoor scenes. Nine scenes from the Gibson dataset were selected for training, and the trained model was then tested on the MatterPort3D dataset. Some scenes with small areas or disconnected regions that were impassable for the TurtleBot robots were excluded, resulting in 51 scenes for performance testing. These scenes were further divided into three subsets based on their area sizes: large, moderate, and small area scenes. During testing, each scene was evaluated over 100 trials, and the average results were recorded. The initial positions and orientations of the robots were randomly generated within the scene, with multiple robots initially concentrated in a small area.\par
\subsubsection{Parameter Settings and Training Details}\par
In this study, the grid map size was set to $480 \times 480$, where each grid cell represents an area of 0.01 square meters in the real world. The maximum field of view radius for the robot was set to 3 meters, and the maximum robot movement speed was set to 1 meter per second. During training, the maximum time steps per episode were set to 1800, and the planning horizon for long-term goal planning was set to 15 time steps. The structure feature encoding network consisted of a 5-layer convolutional neural network, with the output partial-map feature mapping channel set to 32. The feature utilization network was a 3-layer multilayer perceptron network. Additionally, the multilayer perceptron network for encoding node categories and geometric information also had an output layer size of 32. The mutual information evaluation network first encoded the input information using a 4-layer convolutional neural network, followed by a 3-layer multilayer perceptron for feature output, where the output layer size of the perceptron was 64.\par
Furthermore, in the graph matching decision network based on graph attention network, the vector lengths corresponding to the keys, values, and queries in the attention mechanism were set to 32. The code framework used in this study was the widely used PyTorch framework in academia. The asymmetric actor-critic training framework was an improvement based on the Proximal Policy Optimization (PPO) algorithm, with training hyperparameters set as shown in Table I. The above training parameters were determined through experimental comparisons to obtain the optimal values. The code was deployed and trained on a workstation equipped with an Intel i9-13900k central processing unit and an NVIDIA GeForce RTX 4090 graphics card, with the complete training process taking approximately 48 hours.\par
\begin{table}[]
\renewcommand{\arraystretch}{1.2}
\setlength{\tabcolsep}{10pt} 
\caption{Training hyperparameter settings.}
\label{table3}
\resizebox{\linewidth}{!}{
\begin{tabular}{l c}
\toprule
Hyperparameter Name & Value \\ 
\midrule
Training rounds & 1100 \\ 
Learning rate & $1 \times 10^{-5}$ \\ 
Incentive discount factor & 0.99 \\ 
Generalized Advantage Estimation discount factor & 0.95 \\ 
Value loss function coefficient $c_1$ & 3.0 \\ 
Strategy entropy coefficient $c_2$ & 1.0 \\ 
MI evaluation loss initial scaling factor $\hat c_3$ & 1.0 \\
MI evaluation loss growth rate $k$ & 0.1 \\
\bottomrule
\end{tabular}
}
\end{table}
\subsubsection{Evaluation Metrics}\par
For multirobot active mapping tasks in indoor environments, task completion effectiveness is evaluated based on time efficiency and mapping completeness. Time efficiency refers to the time required for robots to complete the exploration task, while mapping completeness reflects the exploration speed within a given time frame. Therefore, time steps and exploration rate are used as evaluation metrics. Time steps indicate the time required for the robots to finish the exploration, while the exploration rate is the ratio of the area explored by the robots to the total explorable area of the environment within the maximum episode length.\par
\subsection{Experimental Results} \par 
\subsubsection{Baseline Methods}
To thoroughly validate the effectiveness of the proposed method, several high-performance baseline methods were introduced for comparison, including four traditional planning methods (Utility \cite{r13}, mTSP \cite{r54}, Voronoi \cite{r46}, CoScan \cite{r11}) and two reinforcement learning-based methods (Ans-Merge \cite{r62}, NCM \cite{r22}). To ensure fair comparison, for the aforementioned baseline methods, only their top-level decision modules, which allocate long-term target points to robots, were utilized. The bottom-level action execution modules for all methods were uniformly processed using a fast traversal algorithm, and the planning horizons for long-term goals were kept the same for all methods. The following provides detailed introductions to the aforementioned baseline methods:\par
\begin{enumerate}[]
\item \textbf{\large Utility} introduces the concept of information gain, where each robot selects the boundary point with the maximum information gain as the long-term target point. The information gain of a boundary point is defined as the area of unexplored regions within a circle centered at that boundary point with the perception distance limit as the radius.\par
\item \textbf{\large mTSP} transforms the multirobot active mapping problem into a multiple Traveling Salesman Problem, which requires multiple robots to cooperatively traverse all boundary point nodes starting from their current node positions. This is achieved by establishing a boundary point-robot passable topological graph containing distance information. \par
\item \textbf{\large Voronoi} segments the entire map using the Voronoi partitioning method, with the robot location as seed points. Each resulting map sub-block ensures that any point within it is closer to its corresponding seed point than to any other seed points. Each robot then selects the nearest boundary point within its map sub-block as the long-term target point.\par
\item \textbf{\large CoScan} first performs K-means clustering on all boundary points and models the multirobot active mapping task as an Optimal Mass Transport Problem, allocating boundary point clusters based on distances between robots and boundary point clusters. \par
\item \textbf{\large Ans-Merge} extends the ANS \cite{r62} method, which is a reinforcement learning-based algorithm for single-robot exploration in unknown environments. It overlays the local grid map centered on itself and the global grid map as decision inputs and selects long-term target points for robots through regression. \par
\item \textbf{\large NCM} builds a topological graph between boundary points and robots based solely on geometric distance information and introduces a multi-graph neural network to predict the neural distance between boundary points and robots. It then matches boundary points with robots based on neural distance and assigns long-term target points to each robot.\par
\end{enumerate}

\begin{table*}[]
\renewcommand{\arraystretch}{1.2}
\setlength{\tabcolsep}{10pt} 
\caption{Performance comparison in MatterPort3D test dataset.}
\label{table3}
\resizebox{\linewidth}{!}
{
\begin{tabular}{l S[table-format=4.2] S[table-format=2.2] S[table-format=4.2] S[table-format=2.2] S[table-format=4.2] S[table-format=2.2]}
\toprule
\multirow{2}{*}{Methods} & \multicolumn{2}{c}{Small Scene ($<60m^2$)} & \multicolumn{2}{c}{Medium Scene ($60-100m^2$)} & \multicolumn{2}{c}{Large Scene ($>100m^2$)} \\
 & {Time (step)} & {Explo (\%)} & {Time (step)} & {Explo (\%)} & {Time (step)} & {Explo (\%)} \\ 
\midrule
Utility\cite{r13} & {1111.84} & {97.28} & {1779.78} & {95.41} & {3056.00} & {83.84} \\
mTSP\cite{r54}   & {893.26}  &  \textbf{97.85} & {1120.22} & {96.76} & {1764.36} & {95.93} \\
Voronoi\cite{r46}  & {904.53}  & {97.68} & {1226.72} & {96.72} & {1520.07} & \textbf{96.19} \\
CoScan\cite{r11}  & {716.63}  & {98.09} & {1070.11} & {96.72} & {1601.43} & {96.07} \\
Ans-Merge\cite{r62} & {1529.58}  & {96.08} & {2425.67} & {85.98} & {3827.21} & {81.19} \\
NCM\cite{r22}    & {690.16}  & {97.78} & {987.44}  & {96.74} & {1492.07} & {96.09} \\
AIM-Mapping      & \textbf{542.35}  & {97.67} & \textbf{803.37}  & \textbf{96.76} & \textbf{1341.59} & {96.03} \\
\bottomrule
\end{tabular}
}
\vspace{-0.3cm}
\end{table*}
\subsubsection{Comparison of Exploration Performance}\par
\begin{figure}
    \centering
    \includegraphics[width=1\linewidth]{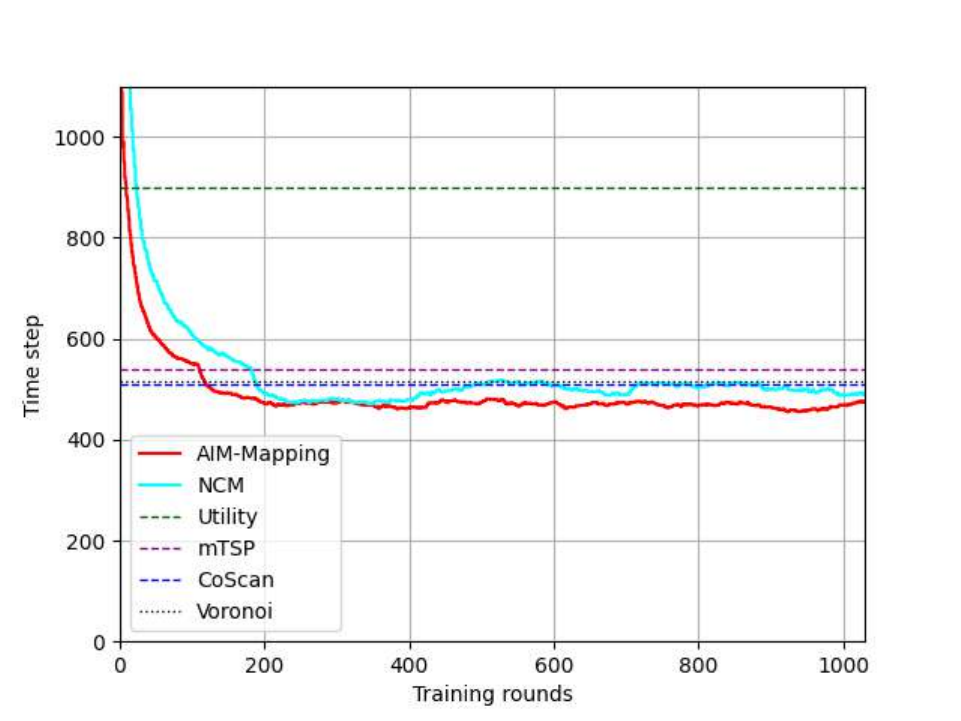}
    \caption{ Training performance comparison. The training results of the proposed AIM-Mapping and the baseline method NCM, as well as the comparison of the performance of the planning-based baseline methods (Utility, mTSP, Voronoi, and CoScan) on the training set.
}
    \label{Training_performance comparison}
\end{figure}
\begin{figure}
    \centering
    \includegraphics[width=1\linewidth]{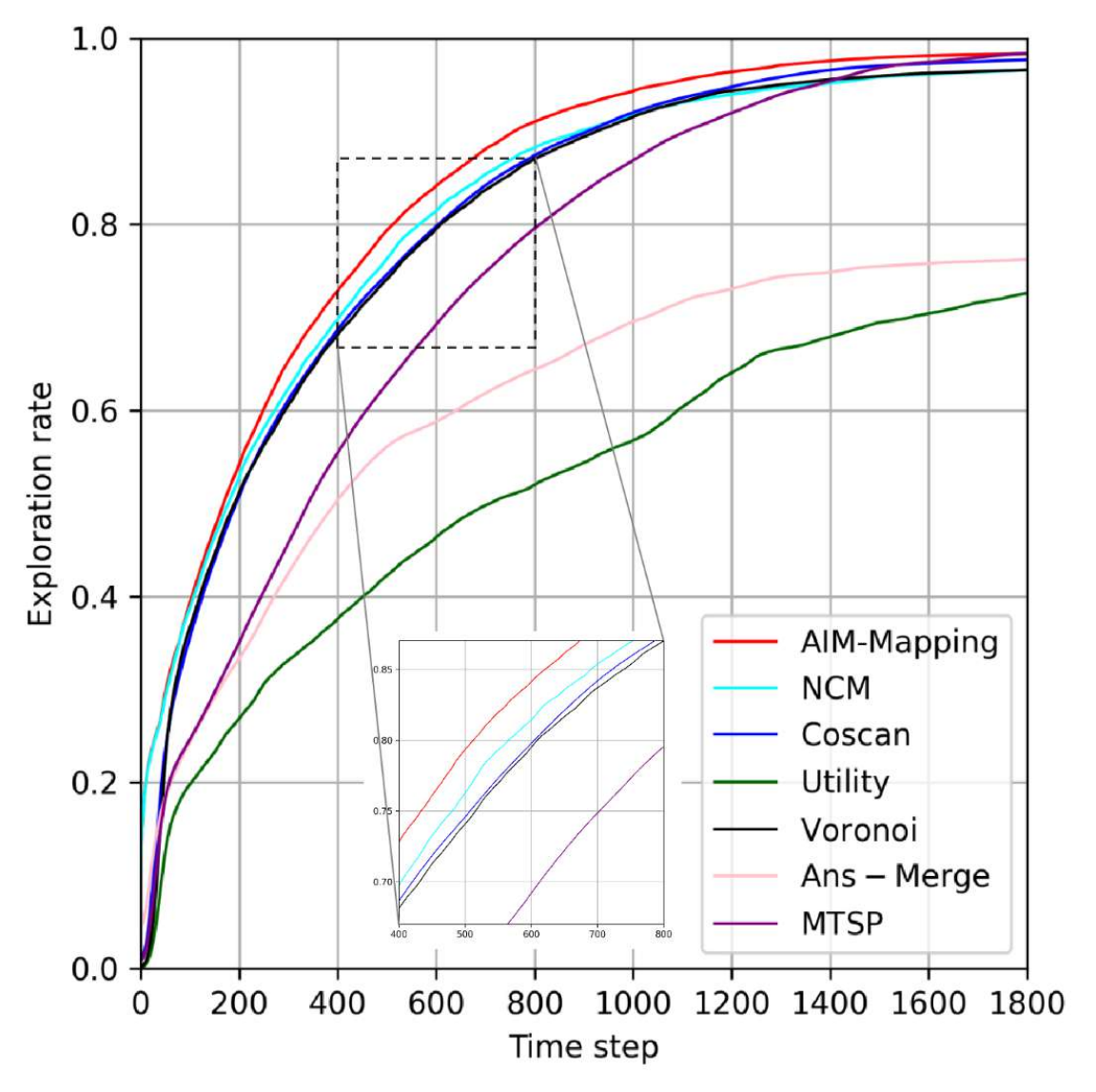}
    \caption{ Illustration of average exploration rate variation during test episodes.
}
    \label{Illustration of average exploration rate variation during test episodes.}
\end{figure}

We recorded the number of time steps required for multiple robots to completely explore the environment during training as the training result. From Fig. \ref{Training_performance comparison}, it can be observed that with the increase in training epochs, the number of time steps required for the AIM-Mapping algorithm and the NCM algorithm to complete the exploration task gradually decreases and eventually outperforms traditional planning methods. Among them, the AIM-Mapping algorithm slightly outperforms the NCM algorithm in terms of convergence speed and performance after convergence. The Voronoi algorithm and the CoScan algorithm perform very similarly on the training set. The average step lengths for completing the exploration task are 486.55 and 488.89, respectively, so the corresponding two dashed lines in Fig. \ref{Illustration of average exploration rate variation during test episodes.} are very close.\par
To further validate the effectiveness of the proposed method, we tested the trained models on the test set and compared them with baseline methods. The comparative experimental results are shown in Table II. From the table, it can be observed that except for the Ans-Merge and Utility methods, all other methods achieve an exploration rate of over 95$\%$ in scenes of various area scales, indicating successful completion of full environment exploration. The Ans-Merge method, using the original grid map as input under the scenes and reward settings of this study, exhibited instability during training, resulting in poor model performance. The Utility method only considers information gain and ignores distance information, leading to significant path redundancy and difficulty achieving high exploration rates in large-scale scenes. However, in terms of time efficiency, the proposed AIM-Mapping outperforms various baseline methods by achieving relatively optimal efficiency at the same exploration rate. In moderate-sized and large-scale scenes, AIM-Mapping reduces the number of time steps required for mapping compared to the best-performing baseline method NCM by approximately 10$\%$. The performance improvement is due to the proposed AIM-Mapping, which not only extracts distance information and structural information from the environment to establish a topological representation but also considers the information value acquired during the map exploration process using mutual information. This results in more effective long-term goal planning, thereby improving the time efficiency of task completion.\par
\subsection{Visualization of Exploration Process}\par
\begin{figure}
    \centering
    \includegraphics[width=0.8\linewidth]{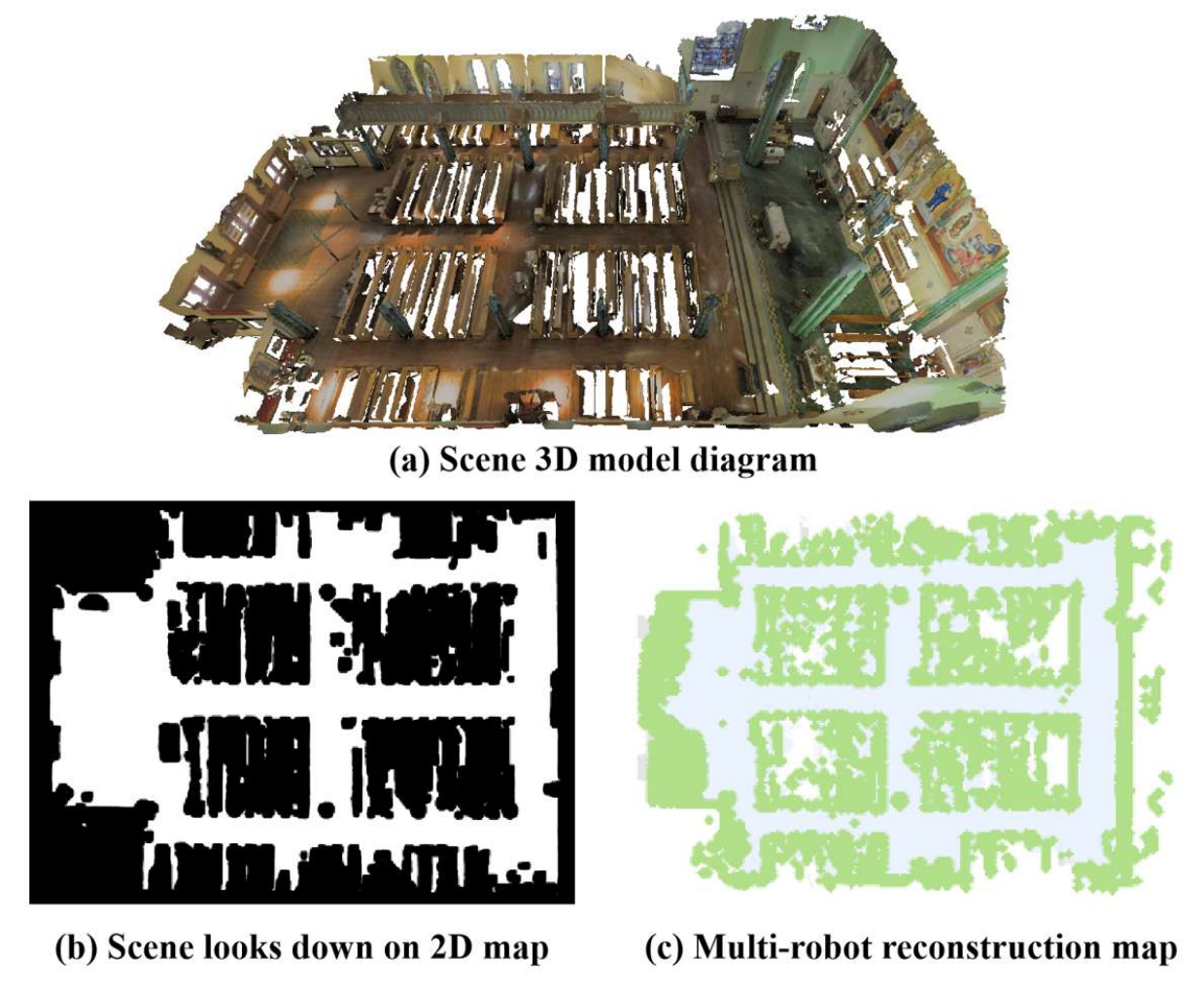}
    \caption{  Visualization and map reconstruction of the scene with the ID 'gYvKGZ5eRqb'
}
    \label{Visualization and map reconstruction effect of the scene with the ID 'gYvKGZ5eRqb'}
\end{figure}
\begin{figure}
    \centering
    \includegraphics[width=1\linewidth]{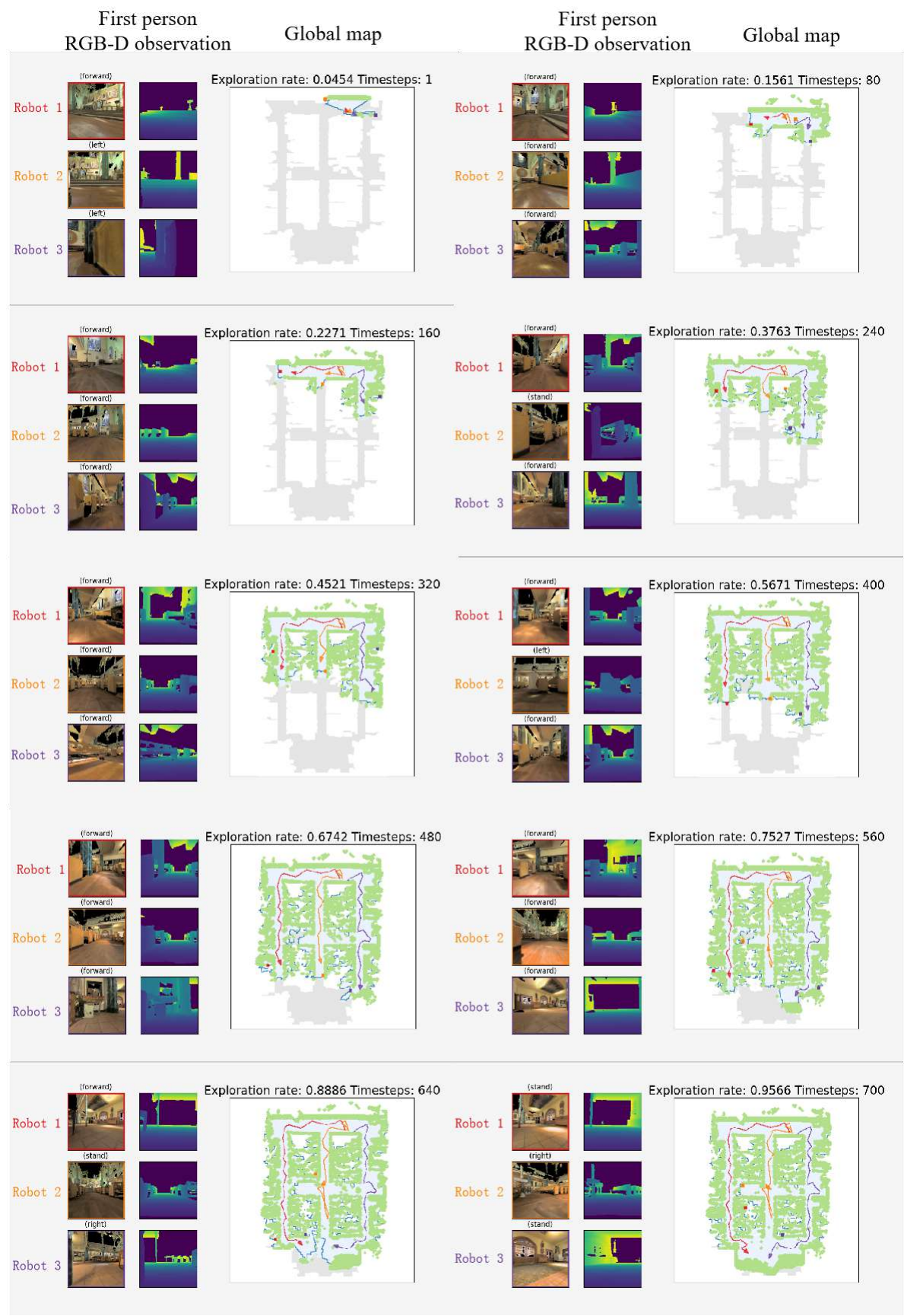}
    \caption{Visualization schematic diagram of robot mapping process.
}
    \label{Visualization schematic diagram of robot mapping process}
\end{figure}

To visually demonstrate the effectiveness of the algorithm, we visualized the testing process in the simulation environment. As shown in Fig. \ref{Visualization and map reconstruction effect of the scene with the ID 'gYvKGZ5eRqb'}, the scenario labeled “gYvKGZ5eRqb” from the test set was used for visualization. Based on the 3D model diagram, the scenario is identified as an indoor auditorium. In this scenario, three robots were deployed using the proposed AIM-Mapping algorithm to reconstruct the map, with the reconstruction result shown in Fig. \ref{Visualization and map reconstruction effect of the scene with the ID 'gYvKGZ5eRqb'}(c). In Fig. \ref{Visualization and map reconstruction effect of the scene with the ID 'gYvKGZ5eRqb'}(c), the green areas represent obstacles, while the light blue areas represent the explored and navigable regions. Since the dataset simulates a real-world scenario, where the ground may be uneven (e.g., protrusions or depressions), there is a slight difference between the left side of the map in Fig. \ref{Visualization and map reconstruction effect of the scene with the ID 'gYvKGZ5eRqb'}(c) and Fig. \ref{Visualization and map reconstruction effect of the scene with the ID 'gYvKGZ5eRqb'}(b). For areas that are inaccessible, the robots mark them as obstacle zones. 
In Fig. \ref{Visualization schematic diagram of robot mapping process}, the paper presents the first-person RGB-D observations of the three robots at different time steps and the global map reconstructed by the multiple robots. In the global map, green areas indicate obstacles, light blue areas indicate explored navigable regions, gray areas represent the true values of navigable areas in the map, and dark blue points denote boundary points. The elements related to the robots are distinguished using different colors: the three robots are represented by red, yellow, and purple, respectively. The arrows in corresponding colors indicate the current positions of the robots, and the curves connected to the arrows show the robots historical trajectories. In the boundary regions between known and unknown maps, dots in corresponding colors represent the long-term goal points assigned to each robot. It can be seen that although the robots were initialized in a small area, they quickly moved in different directions after the exploration began. During the exploration process, their trajectories seldom overlapped, indicating that the area of repeated exploration was minimal and that the long-term decision-making of the robots was efficient.\par
\begin{figure}
    \centering
    \includegraphics[width=1\linewidth]{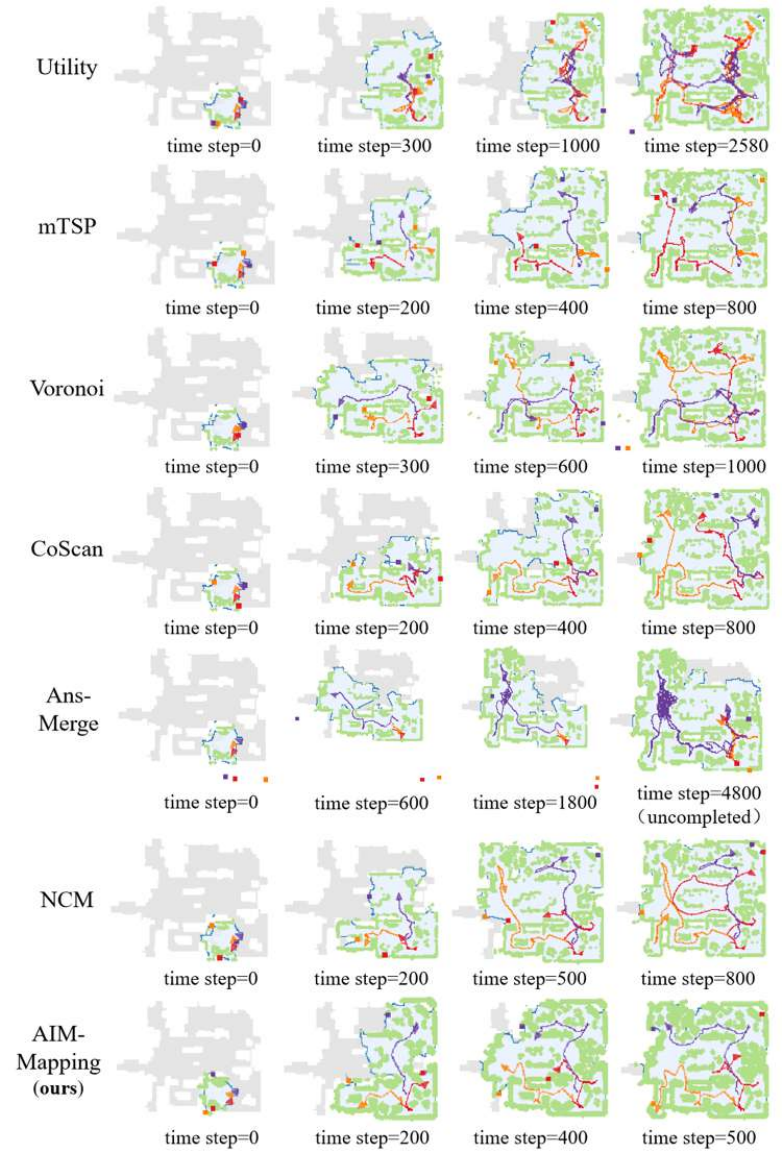}
    \caption{Visualization comparative schematic diagram of mapping process.}
    \label{Visualization comparative schematic diagram of mapping process}
\end{figure}
To provide a clear comparison with various baseline methods, this paper tested each baseline method and recorded the mapping progress and robot trajectories at different time steps during the testing process. Taking the scenario “JmbYfDe2QKZ” as an example, the initial positions and orientations of the robots were the same in each test round, and the visualization results are shown in Fig. \ref{Visualization comparative schematic diagram of mapping process}. The meanings of the colored elements in Fig. \ref{Visualization comparative schematic diagram of mapping process} are the same as those in Fig. \ref{Visualization schematic diagram of robot mapping process}. During the visualization process, we also recorded the time taken by each method to complete the exploration. It can be seen that the AIM-Mapping method achieved better exploration efficiency, completing the overall exploration of the environment in a relatively short time. Additionally, except for Ans-Merge, most of the methods successfully completed the full exploration of the unknown environment and reconstructed a top-down map. This is because Ans-Merge selects long-term goal points using a regression-based approach, choosing a point anywhere on the map as the long-term goal point, without ensuring that the selected point is a boundary point. As a result, Ans-Merge cannot guarantee complete exploration of the environment. Furthermore, under the Utility method, the trajectories of multiple robots show significant overlap, and individual robots exhibit repeated movements, which is consistent with the analysis in Table II.
Finally, among the various methods, the AIM-Mapping method shows less overlap in robot trajectories, and the movements of the robots are relatively smoother. This suggests that the long-term goal point selection of the AIM-Mapping method is more reasonable and efficient to some extent.
\subsection{Ablation and Generalization Experiments}\par
\subsubsection{Generalization Experiment}
To verify the generalization ability of the proposed method with different numbers of robots, the models trained with 3 robots were extended to settings with 4 and 5 robots for testing. The average test results on the entire test set are shown in Table III. From the data in the table, it can be observed that despite the change in the number of robots during testing, the AIM-Mapping method proposed in this paper still achieves relatively superior time efficiency compared to the baseline methods. The strong generalization ability of AIM-Mapping is partly attributed to its construction of the topological representation graph, where topological relationships and distance information are minimally impacted by changes in the number of robots. This indicates that the AIM-Mapping model trained in only 9 scenes also has certain performance limitations in terms of generalization to the number of robots.\par
\begin{table*}[]
\renewcommand{\arraystretch}{1.2}
\setlength{\tabcolsep}{10pt} 
\caption{Comparative experimental performance on generalization of robot quantities.}
\label{table3}
\centering  
\resizebox{\linewidth}{!}{
\begin{tabular}{l S[table-format=4.2] S[table-format=2.2] S[table-format=4.2] S[table-format=2.2] S[table-format=4.2] S[table-format=2.2]}
\toprule
\multirow{2}{*}{Methods} & \multicolumn{2}{c}{Number of robots=3} & \multicolumn{2}{c}{Number of robots=4} & \multicolumn{2}{c}{Number of robots=5} \\
 & {Time (step)} & {Explo (\%)} & {Time (step)} & {Explo (\%)} & {Time (step)} & {Explo (\%)} \\ 
\midrule
Utility\cite{r13}   & {1881.27} & {92.93}  & {1681.73} & {93.69} & {1589.45} & {94.90} \\
mTSP\cite{r54}      & {1212.49} & {96.84}  & {1038.67} & {97.13} &  {873.65} & {96.95} \\
Voronoi\cite{r46}   & {1187.22} & {96.83}  &  {999.14} & {97.17} &  {869.63} & {96.95} \\
CoScan\cite{r11}    & {1084.27} & \textbf{97.05} &  {988.78} & \textbf{97.18} &  {835.63} & {97.00} \\
Ans-Merge\cite{r62} & {2476.57} & {88.43}  & {1681.73} & {93.69} & {1536.73} & {95.00} \\
NCM\cite{r22}       & {1015.22} & {96.87}  &  {868.67} & {97.01} &  {761.78} & {97.00} \\
AIM-Mapping         &  \textbf{763.14} & {97.02}  &  \textbf{747.74} & {97.03} &  \textbf{734.66} & \textbf{97.00} \\
\bottomrule
\end{tabular}
}
\vspace{-0.3cm}
\end{table*}
\subsubsection{Ablation Experiments}\par
\begin{table*}[]
\renewcommand{\arraystretch}{1.2}
\setlength{\tabcolsep}{10pt} 
\caption{Ablation experiment results.}
\label{table3}
\resizebox{\linewidth}{!}{
\begin{tabular}{l S[table-format=4.2] S[table-format=2.2] S[table-format=4.2] S[table-format=2.2] S[table-format=4.2] S[table-format=2.2]}
\toprule
\multirow{2}{*}{Methods} & \multicolumn{2}{c}{Small Scene ($<60m^2$)} & \multicolumn{2}{c}{Medium Scene ($60-100m^2$)} & \multicolumn{2}{c}{Large Scene ($>100m^2$)} \\
 & {Time (step)} & {Explo (\%)} & {Time (step)} & {Explo (\%)} & {Time (step)} & {Explo (\%)} \\ 
\midrule
Ablated privileged & {605.08} & {97.32} & {950.00} & {96.62} & {1495.91} & {96.16} \\
Ablated MI evaluation & {686.70} & {97.55} & {977.90} & \textbf{96.82} & {1481.36} & \textbf{96.20} \\
AIM-Mapping &\textbf{542.35} & \textbf{97.67} & \textbf{803.37} & {96.76} & \textbf{1341.59} & {96.03} \\
\bottomrule
\end{tabular}
}
\vspace{-0.3cm}
\end{table*}
To further validate the effectiveness of each module proposed in the AIM-Mapping method, we present the results of ablation experiments on different modules. Ablated privileged representation: The privileged information introduced in the asymmetric feature representation module is removed, and only current and historical observations are used to evaluate the state value. Ablated mutual information evaluation: The mutual information evaluation module, which assesses environmental uncertainty using privileged information, is removed, and the robots are guided using the explored area for exploration. The comparison test results of the trained models on the test set are presented in Table IV.\par

Experimental results show that removing privileged information reduces time efficiency of the algorithm, as reflected in the tests conducted on three scenarios of different sizes. This, to some extent, indicates that utilizing privileged global information in the network evaluation module during training not only helps the feature encoding module capture accurate and valuable feature information but also allows for a more accurate assessment of the robots state and action values, thereby improving decision-making efficiency. Experimental results on the test set also indicate that not using privileged information reduces the robots decision-making efficiency. Thus, using global information as privileged input, extracting feature mappings through asymmetric feature representation, and evaluating map uncertainty during exploration with the mutual information evaluation method can improve the overall decision-making performance of the system.\par

\subsection{Real-world Experiments}\par
To further validate the effectiveness of AIM-Mapping, we designed the real-world experiments.\par
\begin{figure}
    \centering
\includegraphics[width=1 \linewidth]{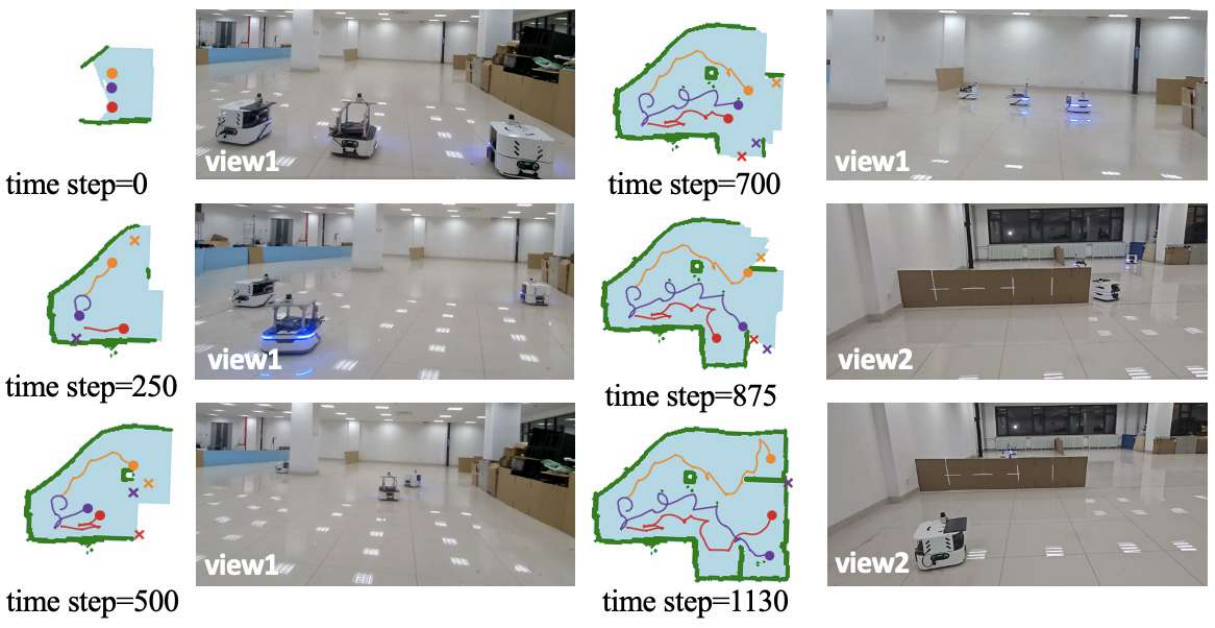}
    \caption{An illustration of multirobot exploration in a real-world scene. The exploration process is recorded using two cameras from different view points.
}
    \label{An illu real}
\end{figure}

As shown in Fig. \ref{An illu real}, in our real-world experiments, we use three TRACER MINI AGVs of AgileX Robotics, and each of them is mounted an H1 Series LiDAR of Free Optics as the depth sensor.  
In addition, each robot is equipped with an onboard processing unit featuring an Intel Core i5-8265U CPU @ 1.60 GHz (14 nm process) with 4 cores and 8 threads. For centralized processing, we use a laptop configured with an AMD Ryzen 7 7735H processor with Radeon Graphics @ 3.20 GHz, along with an NVIDIA GeForce RTX 4060 GPU, serving as the centralized server.
Specifically, the server receives the partial-maps and localization data from each robot, integrates them to generate a joint map, and then transmits the updated joint map back to each robot for decision-making. Such information transmission is achieved through a local area network.
We directly use the policy model trained in the simulated environments for robots in real-world experiments.
There are two real-world scenes, and the areas of them are both larger than $100 m^2$. 
For each scene, we run AIM-Mapping and baseline methods three times and use the average exploration time and rate for evaluation.\par

\begin{figure}
    \centering
\includegraphics[width=1 \linewidth]{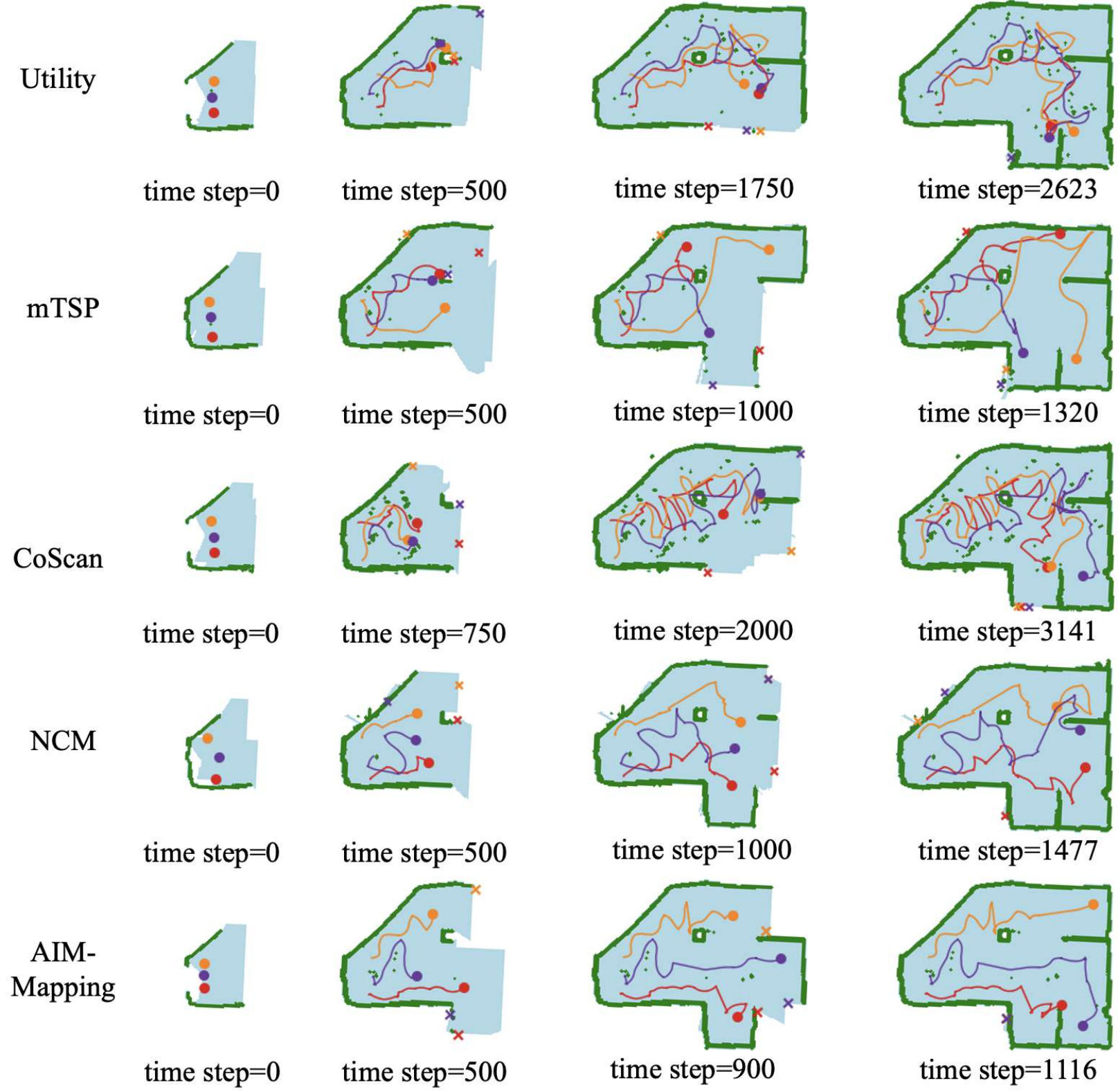}
    \caption{A qualitative comparison of AIM-Mapping and baseline methods in Scene 1. The trajectories of three robots are denoted as three lines with different colors.
}
    \label{comparison1.}
\end{figure}

\begin{figure}
    \centering
\includegraphics[width=1 \linewidth]{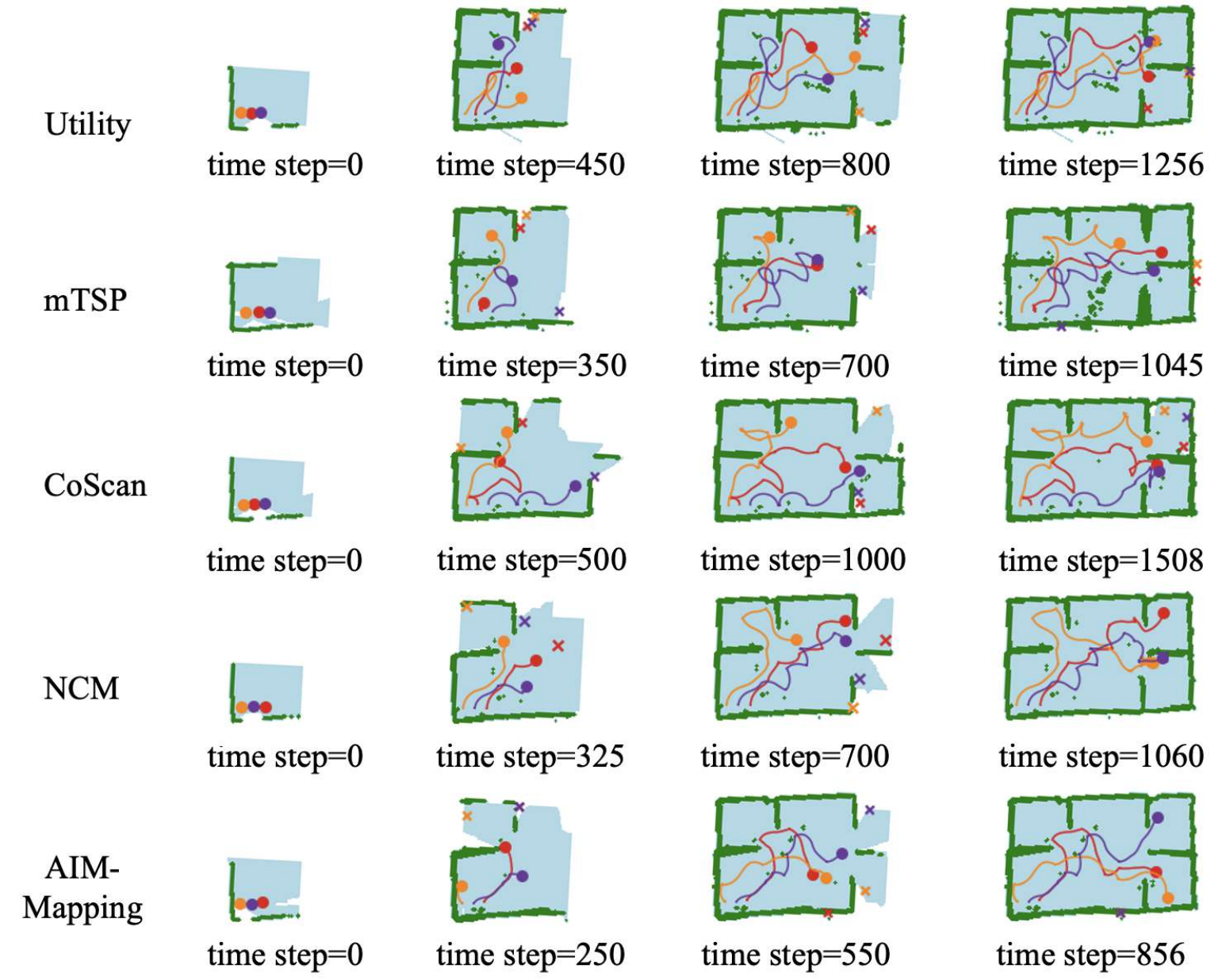}
    \caption{A qualitative comparison of AIM-Mapping and baseline methods in Scene 2. The trajectories of three robots are denoted as three lines with different colors.
}
    \label{comparison2.}
\end{figure}

Fig. \ref{comparison1.} and Fig. \ref{comparison2.} show the qualitative comparison of AIM-Mapping and baseline methods in two real-world scenes. Compared with other methods, with the same time consumption, AIM-Mapping explored more area, and firstly finished the exploration task. Aligned with the best exploration performance, there are less overlaps of the trajectories for three robots, demonstrating that the multirobot system can make well task assignment and reduce redundant exploration.\par
Table V shows the quantitative comparison in two real-world scenes. Consistent with the analysis in Fig. \ref{comparison1.} and Fig. \ref{comparison2.}, AIM-Mapping achieves the best performance in terms of exploration time. In the two scenes, all of the methods succeed in exploring the whole area, this is because in the real-world scenes, there are no regions that robots cannot get into or perceive from any viewpoint. Note that, AIM-Mapping with policy network trained in simulation environments achieves superior performance to other methods in both simulation and real-world scenarios, which demonstrate the effectiveness and generalization.

\begin{table}[]
\renewcommand{\arraystretch}{1.2}
\setlength{\tabcolsep}{1pt} 
\caption{Quantitative comparison in two real-world scenes.}
\label{realtab}
\resizebox{\linewidth}{!}
{
\begin{tabular}{l S[table-format=4.2] S[table-format=2.2] S[table-format=4.2] S[table-format=2.2] S[table-format=4.2] S[table-format=2.2]}
\toprule
\multirow{2}{*}{Methods} & \multicolumn{2}{c}{Scene 1} & \multicolumn{2}{c}{Scene 2}  \\
 & {Time (step)} & {Explo (\%)} & {Time (step)} & {Explo (\%)} \\ 
\midrule
Utility\cite{r13} & {2613.67} & {100.00} & {1255.33} & {100.00} \\
mTSP\cite{r54}   & {1322.00}  &  {100.00} & {1095.00} & {100.00} & \\
CoScan\cite{r11}  & {3148.67}  & {100.00} & {1522.33} & {100.00} \\
NCM\cite{r22}    & {1462.33}  & {100.00} & {1053.33}  & {100.00} \\
AIM-Mapping      & \textbf{1111.67}  & \textbf{100.00} & \textbf{851.67}  & \textbf{100.00} & \\
\bottomrule
\end{tabular}
}
\vspace{-0.3cm}
\end{table}
\subsection{Discussion}\par
Despite achieving superior performance compared to existing methods, AIM-Mapping has several limitations.
Firstly, the decision-making process primarily relies on 2D map information converted from RGB images and depth data, which can result in information loss, particularly in terms of semantic information.
Secondly, in AIM-Mapping, robots share a common joint map, which is updated based on the observations from each robot. This process assumes no communication delays or failures, which may not hold true in more complex environments.
Thirdly, the sensors used in our experiments are either noise-free or experience minimal noise. In contrast, real-world scenarios, especially in dynamic environments, may involve significant disturbances that could degrade the exploration performance of AIM-Mapping. What's more, there are three sub-task modules in multirobot exploration, and AIM-Mapping primarily focuses on the long-term goal selection. For perception and short-term path planning, more research effort is needed to investigate the effect of them on the task especially in challenging scenarios like dynamic environments. In the future work, we will further address the above-mentioned issues, and improve the efficiency and robustness of the whole system.

In addition, AIM-Mapping benefits significantly from the introduction of asymmetric information. Incorporating more intrinsic or auxiliary information may further enhance exploration efficiency. 
With additional information, such as semantic data, robots will gain a better understanding of the environment, offering new insights.
As some existing works have demonstrated, map prediction offers explicit information gain that can guide exploration more effectively. In future work, we plan to investigate the impact of map prediction on multirobot exploration and explore its integration with AIM-Mapping to achieve improved performance. Also, further work should be carried out in more complex scenarios, such as multi-storied indoor environments and outdoor field settings. In these challenging conditions, robots often face issues such as perception robustness, communication delays, and increased coordination demands, all of which require more advanced collaboration strategies.

\section{Conclusion}
\label{Conclusion}
This paper studies the multirobot active mapping problem and proposes AIM-Mapping, which is an effective mapping framework based on deep reinforcement learning. 
The framework uses an asymmetric feature representation module to encode the disparity between partial-map and privileged information, and use the disparity feature as the state value of the actor-critic training framework. The mutual information between partial-map and privileged information will be used as the supervised information of the above framework. For decision-making, a topological representation is first constructed incorporating both structural information and geometric distance information. A graph matching mechanism is then applied to assign the goal point to each robot. Qualitative and quantitative experiments are conducted on both the public iGibson environment and real-world scenarios, and the results validate the effectiveness of the proposed method. In our future work, we plan to further explore the potential of the asymmetric information.
\par

\printbibliography

\vspace{-0.5cm}
\begin{IEEEbiography}[{\includegraphics[width=1in,height=1.25in,clip,keepaspectratio]{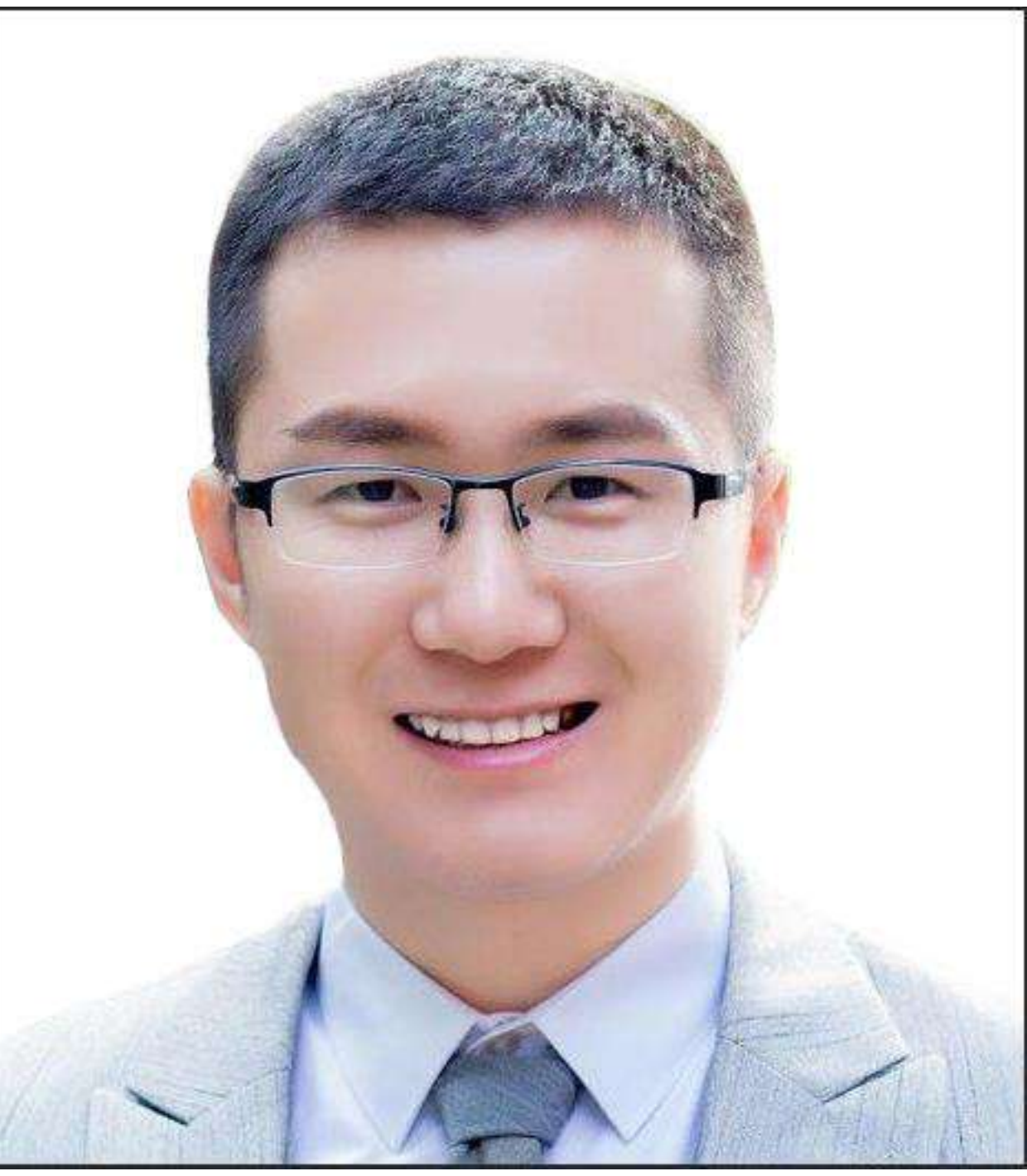}}]{Jiyu Cheng}
received the B.E. degree in automation from Shandong University, Jinan, China, in 2015, and the Ph.D. degree in electronic engineering from The Chinese University of Hong Kong, Hong Kong, in 2019. He is currently an Associate Professor with the Department of Control Science and Engineering, Shandong University. His current research interests include multirobot system, robot perception and skill learning.
\end{IEEEbiography}

\vspace{-0.5cm}
\begin{IEEEbiography}[{\includegraphics[width=1in,height=1.25in,clip,keepaspectratio]{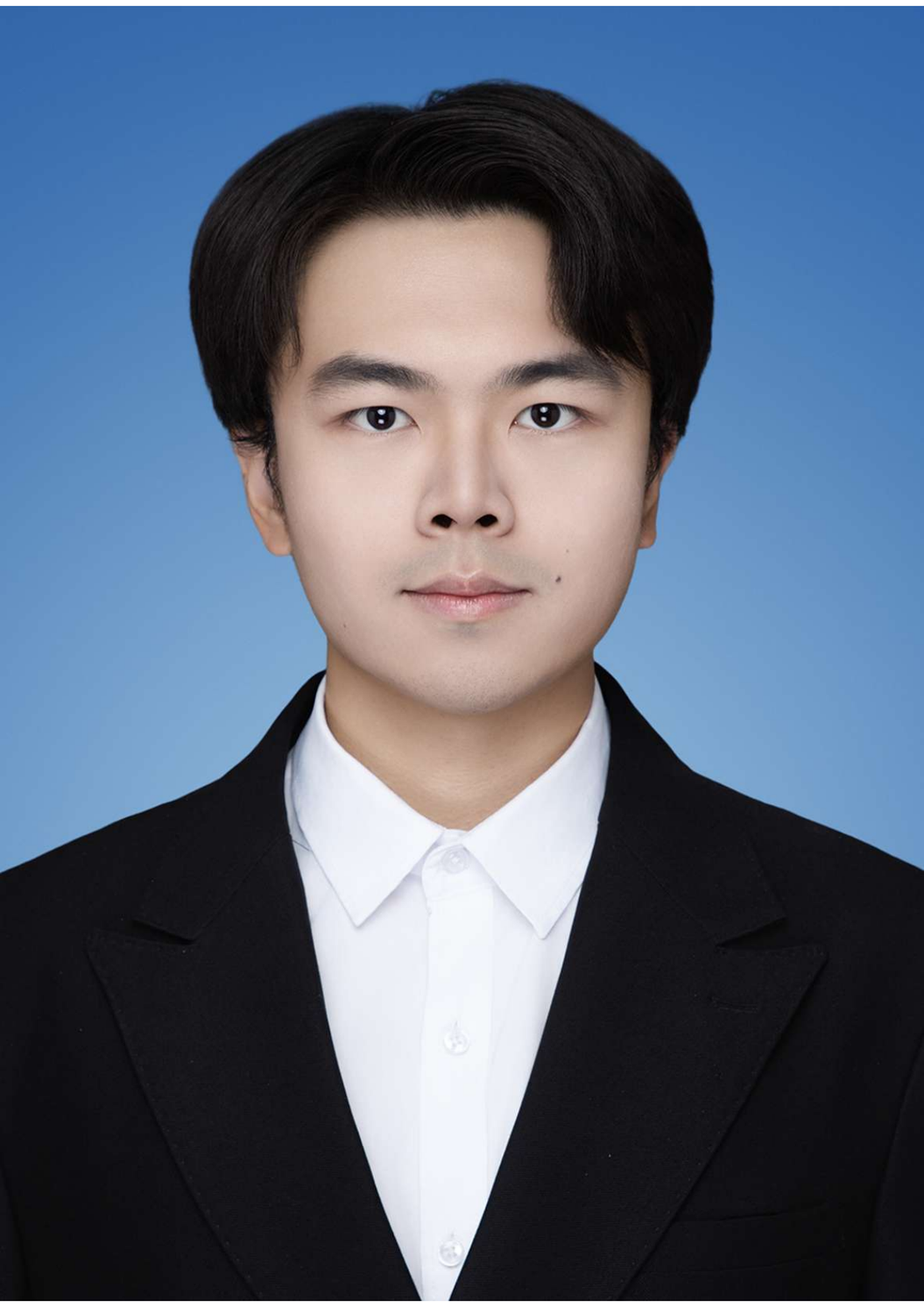}}]{Junhui Fan}
is a graduate student in the School of Control Science and Engineering at Shandong University. He received his B.E. degree in Automation from China University of Mining and Technology. His research interests include multi-robot systems, multi-agent reinforcement learning.
\end{IEEEbiography}

\vspace{-0.5cm}
\begin{IEEEbiography}[{\includegraphics[width=1in,height=1.25in,clip,keepaspectratio]{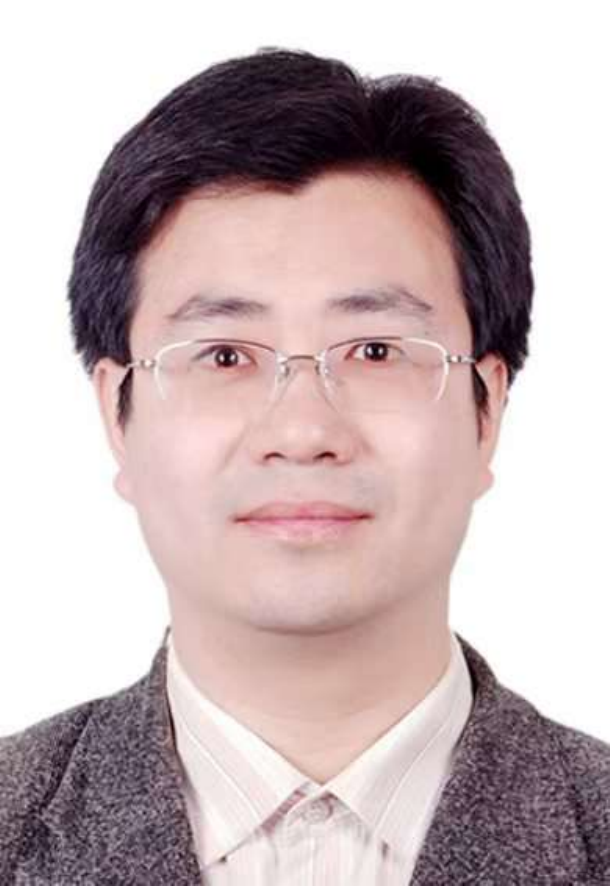}}]{Xiaolei Li}
is currently an Associate Professor with Shandong University, Jinan, China. His research interests include autonomous robots, deep learning, and motion planning.
\end{IEEEbiography}

\begin{IEEEbiography}
[{\includegraphics[width=1in,height=1.25in,clip,keepaspectratio]{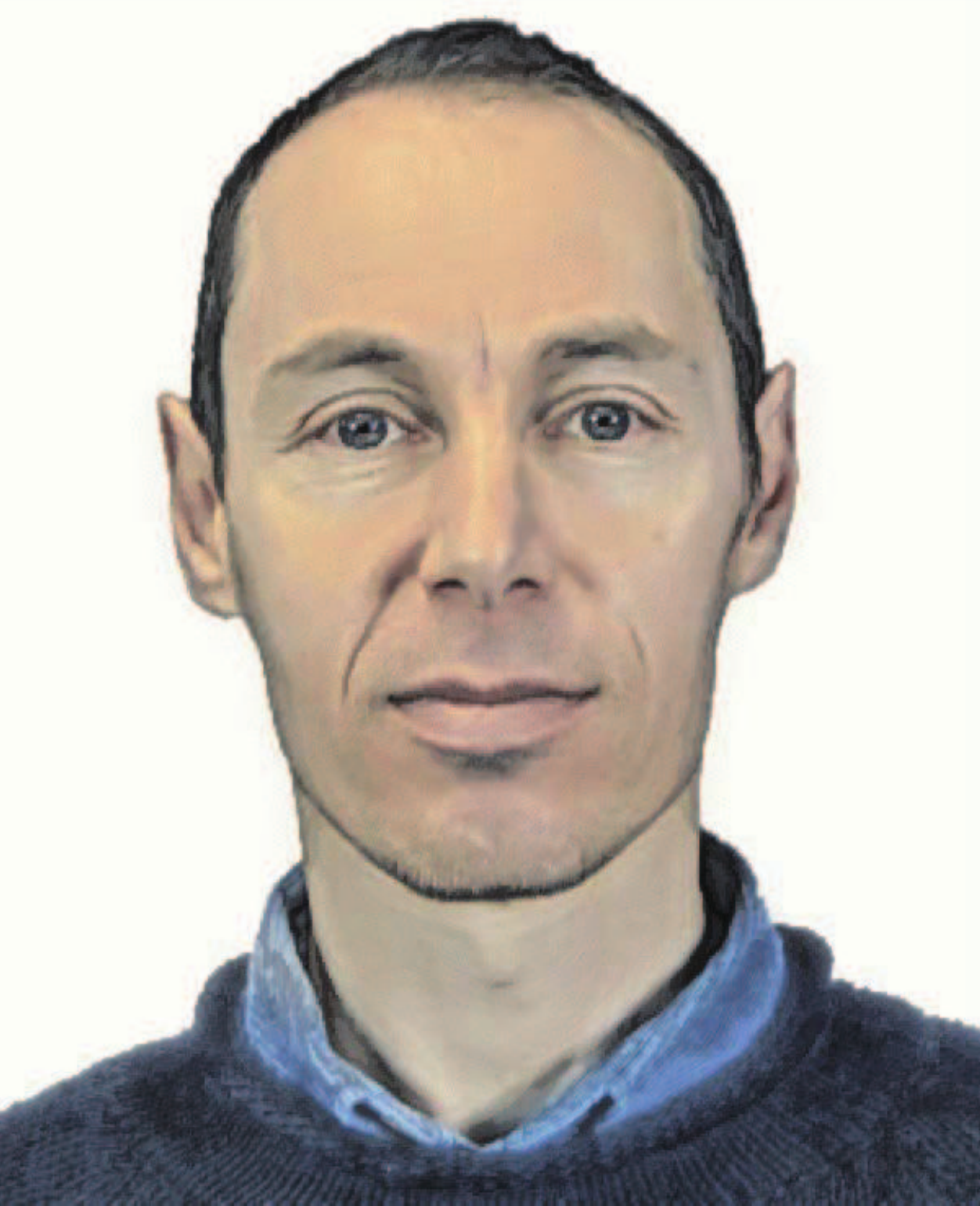}}]{Paul L. Rosin} is a professor with the School of Computer Science and Informatics, Cardiff University, UK. Previously, he was a lecturer at Brunel University
London, U.K., research scientist at the Institute for Remote Sensing Applications, Joint Research Centre, Ispra, Italy, and lecturer at Curtin University of Technology, Perth, Australia. 
His research interests include image representation, semantic segmentation, low level image processing, machine vision approaches to remote sensing, methods for evaluation of approximation algorithms, medical and biological image analysis, mesh processing, non-photorealistic rendering, and analysis of shape in art and architecture.
\end{IEEEbiography}

\begin{IEEEbiography}
[{\includegraphics[width=1in,height=1.25in,clip,keepaspectratio]{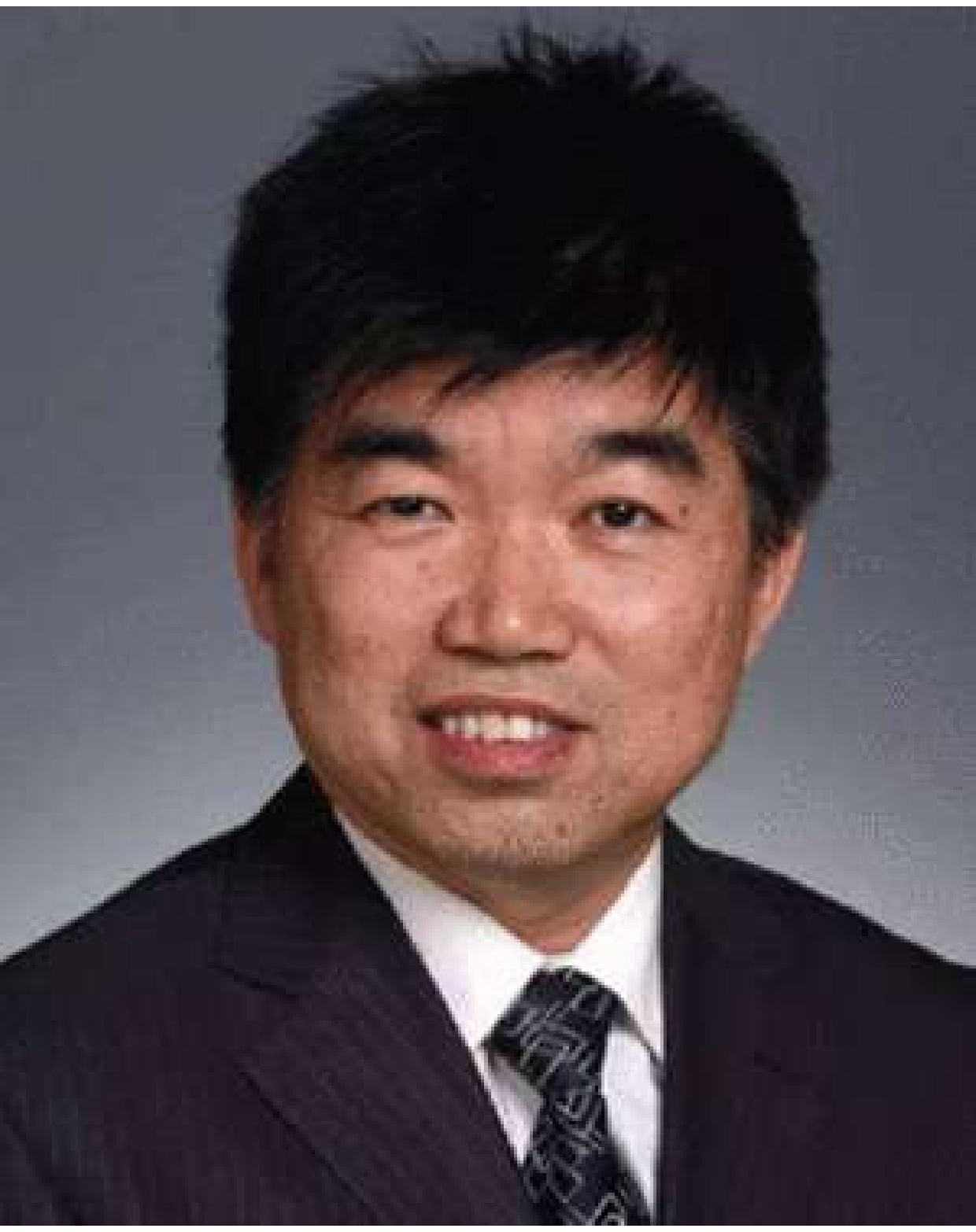}}]{Yibin Li}
received the B.Eng. degree in automation from Tianjin University, Tianjin, China, in 1982, the M.Eng. degree in electrical automation from the Shandong University of Science and Technology, Qingdao, China, in 1990, and the Ph.D. degree in automation from Tianjin University, Beijing, China, in 2008.
From 1982 to 2003, he was with the Shandong University of Science and Technology. Since 2003, he has been the Director of the Center for Robotics, Shandong University, Jinan, China. His research interests include robotics, intelligent control theories, and computer control systems.
\end{IEEEbiography}

\begin{IEEEbiography}
[{\includegraphics[width=1in,height=1.25in,clip,keepaspectratio]{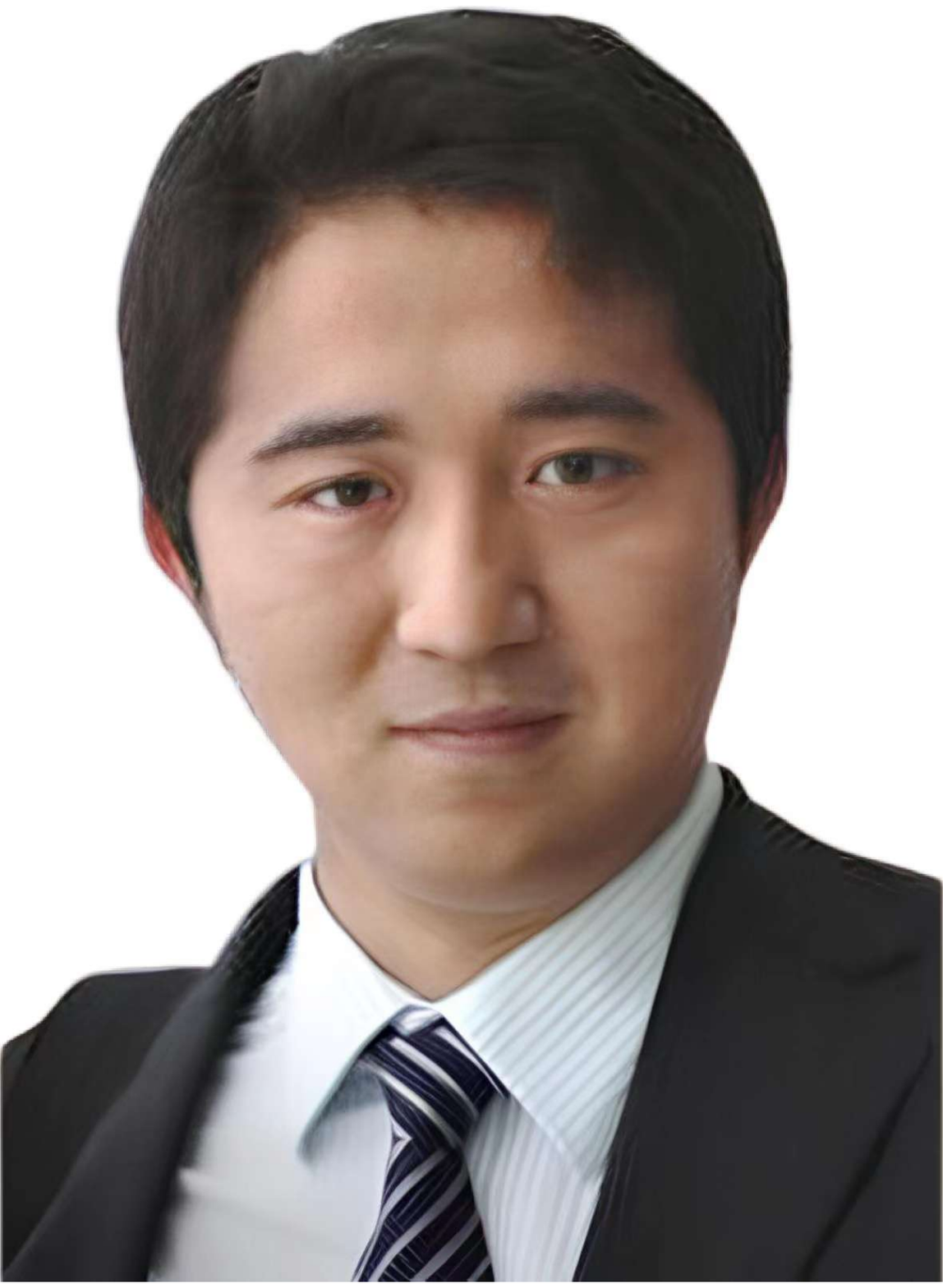}}]{Wei Zhang}
received the Ph.D. degree in electronic engineering from the Chinese University of Hong Kong in 2010. He is currently a professor with the School of Control Science and Engineering, Shandong University, Jinan, China. His research interests include computer vision and robotics. Prof. Zhang has served as a program committee member and a reviewer for various international conferences and journals.
\end{IEEEbiography}

\end{document}